\documentclass{aa}
\usepackage{graphicx,psfig}
\usepackage{natbib}
\usepackage{txfonts}
\def\be{\begin{equation}}
\def\ee{\end{equation}}
\begin{document}
\title{A new code to study structures in collisionally active, perturbed debris discs. Application to binaries.}

\author{P. Thebault\inst{1}}
\institute{
LESIA-Observatoire de Paris, CNRS, UPMC Univ. Paris 06, Univ. Paris-Diderot, France
}

\offprints{P. Thebault} \mail{philippe.thebault@obspm.fr}
\date{Received; accepted} \titlerunning{structures in debris discs}
\authorrunning{Thebault}

\abstract
%
{Debris discs are traditionally studied using two distinct types of numerical models: statistical particle-in-a-box codes to study their collisional and size distribution evolution, and dynamical N-body models to study their spatial structure. The absence of collisions from N-body codes is in particular a major shortcoming, as collisional processes are expected to significantly alter the results obtained from pure N-body runs.
}
%
{We present a new numerical model, to study the spatial structure of perturbed debris discs at dynamical $and$ collisional steady-state. We focus on the competing effects between gravitational perturbations by a massive body (planet or star), collisional production of small grains, and radiation pressure placing these grains in possibly dynamically unstable regions.}
{We consider a disc of parent bodies at dynamical steady-state, from which small radiation-pressure-affected grains are released in a series of runs, each corresponding to a different orbital position of the perturber, where particles are assigned a collisional destruction probability. These collisional runs produce successive position maps that are then recombined, following a complex procedure, to produce surface density profiles for each orbital position of the perturbing body.}
{We apply our code to the case of a circumprimary disc in a binary. We find pronounced structures inside and outside the dynamical stability regions. For low $e_B$, the disc's structure is time varying, with spiral arms in the dynamically "forbidden" region precessing with the companion star. For high $e_B$, the disc is strongly asymmetric but time invariant, with a pronounced density drop in the binary's periastron direction.} {}
\maketitle

\section{Introduction} \label{intro}

Debris discs have been detected around a significant fraction (between $\sim5$ and $\sim30\%$) of main sequence stars of all spectral types \citep[e.g.][]{su06,tril08}. They consist of collisionally interacting solid bodies spanning a wide size range, from $\sim$1-100 kilometre-sized parent bodies, probably leftovers from the planet formation process, down to micron-sized debris. As they make up most of the geometrical cross section, only the smallest, $\leq1\,$cm, particles of this collisional cascade are detectable, in thermal emission or scattered light \citep[see review by][]{wyatt08}.

For most discs that have been resolved, pronounced structures have been observed, such as two-sided asymmetries, spirals, warps, clumps or rings \citep[e.g][]{kala05,goli06,schn09}.
Such features indicate complex dynamical environments, and numerous studies have been aimed at explaining their origin \citep[e.g.][]{kriv10}. Most of these studies are based on $N$-body numerical codes that follow the dynamical evolution of the system, exploring, depending on each system's specificities, different scenarios: perturbing influence of (often undetected) planets or stellar companions, violent transient events, or interaction with gas remnants. In such a collisionless $N$-body approach, the disc is sampled by a population of $N_{num}$ test particles whose trajectories can be precisely tracked. The forces that control their evolution being: the central star's gravity, the gravitational pull of planets and other stars, radiation pressure, Poynting-Robertson (PR) drag, stellar wind drag. The precision of the $N$-body method allows one to study fine effects such as orbital resonances or transient spatial inhomogeneities. However, this comes at a price, which is that collisions are neglected.
This leads to several shortcomings, which can be more or less worrying depending on the issues that are being addressed. These shortcomings can be the following:

\begin{itemize}
\item {\it 1) No size distribution evolution}. Since collisions control the particle size
distribution and its evolution (by merger, erosion or shattering), collisionless N-body codes cannot handle this important issue. In particular, they cannot assess to what extent the specific dynamics of a given system can affect the physical evolution of its population, since the number, size distribution and velocity spread of collisional fragments directly depend on impact velocities, and thus how size distributions can vary as a function of spatial location.
\item {\it 2) No feedback of collisions on the dynamics}. Impacts, be it accreting, bouncing or fragmenting ones, necessarily alter the orbits of colliding objects, because angular momentum is redistributed and kinetic energy is lost to heat. These effects can significantly change the results of pure N-body simulations by damping high eccentricities induced by a planetary or stellar perturbations, or by ejecting collisional fragments far from their progenitors.
\item {\it 3) Particle lifetimes / Spatial structures}. Dynamical processes have characteristic timescales, which can be relatively long, as for instance for mean motion resonances or secular effects. Depending on the disc's density, collision timescales could be shorter than the dynamical ones, thus interfering with, or even preventing the development of the corresponding spatial structures. This problem gets even more complex because collisional timescales are in most cases size-dependent. For the smallest grains, of specific interest because they usually dominate the disc luminosity \citep{theb07}, the dynamical timescales also become size-dependent because of the effect of radiation pressure. These competing timescale effects, in addition to affecting the development of spatial structures, can have direct consequences on the size distribution, which can, at some locations in the disc, strongly depart from any classical equilibrium power-law.
\end{itemize}

Incorporating collisions into an N-body scheme is extremely difficult in the context of debris discs, where typical impact velocities are high and lead to destructive collisions producing a vast amount of small fragments. Following all these fragments would lead to an exponential increase in the number of particles that would soon become unmanageable.

The collisional evolution of debris discs is usually investigated separately with a different type of model, based on statistical particle-in-a-box codes, where the solid body population is distributed into logarithmically-spaced size bins, whose number density is followed using detailed prescriptions for collision outcomes (fragmentation, cratering, accretion). The price to pay here is a poor spatial resolution and a very limited modelling of the dynamics: collision rates are estimated using spatially averaged estimates of orbital parameters that are fixed or follow very simplified evolution laws \citep[e.g.][]{kriv06,theb07}.
A pure statistical approach thus cannot give any precise information on the creation and evolution of spatial structures. For this, an N-body scheme is unavoidable.

\subsection{N-body and collisions: Previous studies}

Given the sheer difficulty of the task, there have only been a few attempts at incorporating fragmenting collisions into N-body codes. 
The conceptually most natural, but in practice most difficult way to do this is by a "brute force" method that follows the fragments produced after each impact. The pioneering study of that method is that of \citet{beau90}, whose code followed 4 fragments after each shattering collision, but had to be restricted to only 200 initial bodies, and was limited to relatively short timescales before reaching a critical number of particles. More recently, \citet{lein05} considered the breakup of rubble piles of gravitationally bound hard spheres, but did not get below the size of the initially defined "hard" sphere units. 

Another solution is to use a mix of the N-body and the statistical approaches. To our knowledge, the only published attempt in this direction is that of \citet{grig07}, using a population of "super particles" (SP), whose orbits are deterministically followed, each representing a cloud of monosize grains. When SPs' paths cross, collisions are considered between their respective grain populations in a statistical way, and new SPs are created to account for the fragments. However, this model was restricted to very short timescales of a few orbital periods. This combined N-body/statistical approach is thus far from being able to address long term phenomenon, but it is probably the most promising one in terms of its potentialities in solving all 3 main problems listed in the previous section.

As of today, the best available model is probably that of \citet{star09}, whose main ambition is to address shortcoming number 3 \footnote{Although not implemented yet, some degree of fragmentation is in principle possible to incorporate in this algoritm \citep{star09}, but probably not enough to fully address points number 1\& 2, especially the way impact velocities locally affect size distributions and velocity spreads by controlling the production of clouds of small fragments} regarding the competing effects of dynamical and collisional lifetimes on the development of spatial structures. Its principle is to release a population of collisionless test particles and record, at regularly spaced time intervals, their position to construct a stream of positions and velocities for each particle. All streams are then recombined to create a surface density map on which the same particles are released once again and removed following collision probabilities derived from the density map. A new map of streams is then created and the process is iterated until it converges. This pioneering model has given impressive results for the specific case of the Kuiper Belt \citep{kuch10}.
In its current version, this code is designed for the specific case of one perturber on a circular orbit. Although the case of multi-perturbers and of an eccentric perturber could in principle be implemented, these issues have not been addressed yet and the code might not be able to handle fast evolving spatial structures (Stark, private communication).
In the case of an eccentric perturber, this code requires that the time between two consecutive records must equal a multiple of the orbital timescale, which must be less than the collision timescale.  As a result, this code may be poor at modelling highly collisional discs with distant perturbers on eccentric orbits, such as the case of a circumprimary debris disc in a binary.

We present here a new code, also aimed at studying how collisional lifetimes affect the development of perturbed spatial structures (issue number 3), which can be used for the generic case of a collisional disc perturbed by one massive body, planet or stellar companion, having any possible orbit, circular or eccentric, internal to, embedded in, or external to the disc. This code can handle size-dependent effects, in particular for small grains placed on high-eccentricity orbits by radiation pressure. The only required assumption is that a dynamical and collisional steady state has been reached in the system. We illustrate this new model with the specific case of a circumprimary debris disc perturbed by an external stellar companion.

\section{Model} \label{model}

\begin{figure*}
\makebox[\textwidth]{
\includegraphics[scale=0.88]{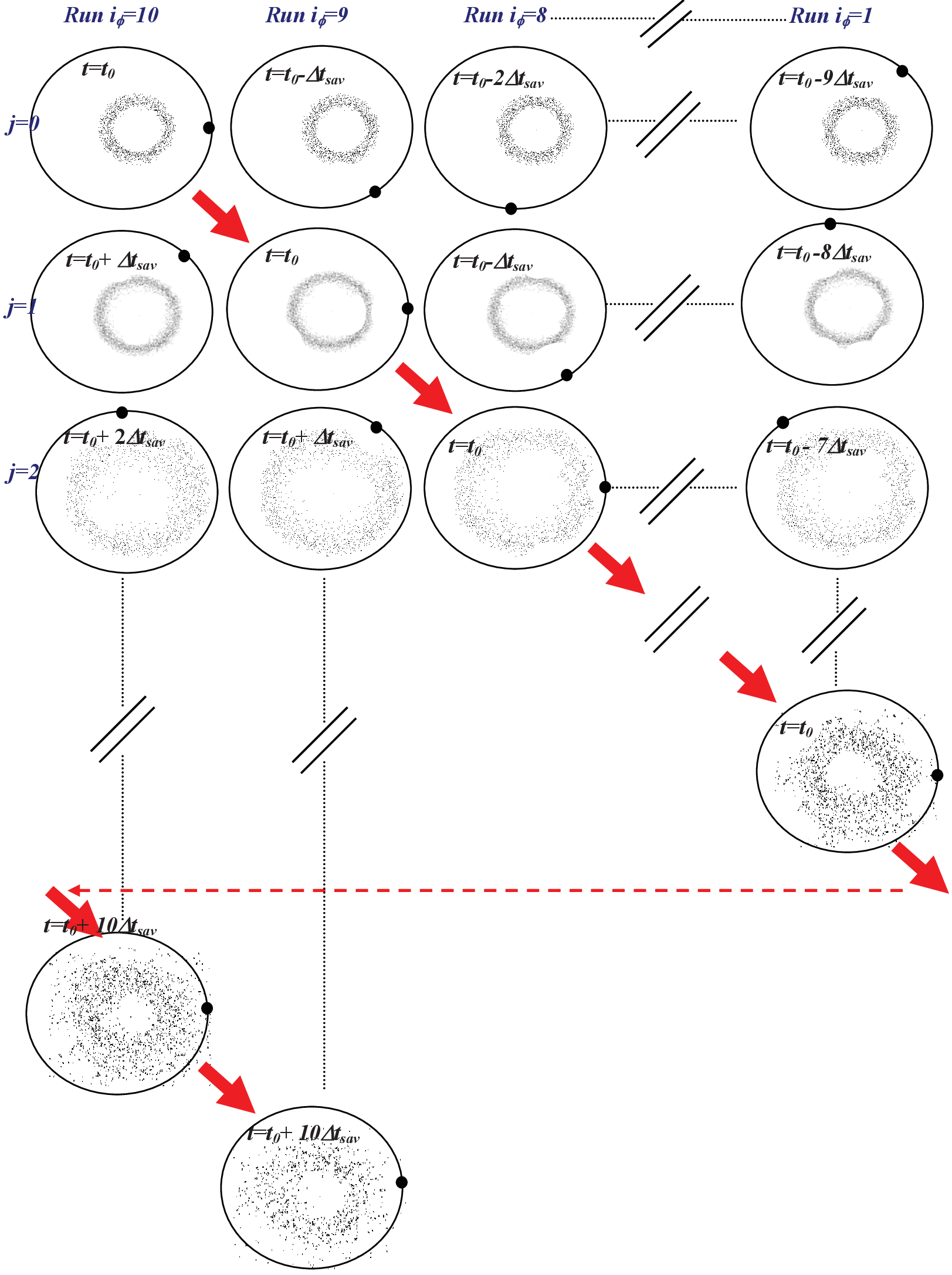}
}

\caption[]{Schematic presentation of the numerical model. $n_{sav}=10$ "collisional" runs are performed, each taking as a starting point one of the $n_{sav}$ parent body (PB) disc configurations, corresponding to $n_{sav}$ different positions of the companion star on its orbit separated by a constant time interval $\Delta t_{sav} = t_{orb}/n_{sav}$, obtained at the end (i.e., steady state) of the parent body run. $N_{num}=2\times10^{5}$ particles are released from the PB's positions, following a size distribution $dN \propto s^{q}ds$ down to the radiation pressure blow out size $s_{RP}$. Particles positions are recorded at time intervals separated by $\Delta t_{sav}$. Each simulation is run until all particles have been removed, either by dynamical encounter with the companion star or by collisions. The collected data is then used to reconstruct the dust distribution at dynamical and collisional steady state by means of the following procedure (indicated by the red arrows on the graph): the dust distribution at a given time $t_0$, corresponding to a given position $i_\phi$ of the companion star, is the combination of the dust released at $t_0$ for the $i_\phi$ run, plus the dust particles released at $t_0-\Delta t_{sav}$ for the $i_\phi -1$ run that have not been removed at $t_0$, plus the dust released at $t_0-2\Delta t_{sav}$ for the $i_\phi -2$ run that has not been removed at $t_0$, etc. The procedure is iterated until we reach the $j_{final}$ record (particles released at $t_0-j_{final}\Delta t_{sav}$), for which no particle has survived until $t_0$. Given that the binary orbit is divided into $n_{sav}=10$ positions, all $i_\phi -10\times j$ runs correspond to the $i_\phi$ one. (The different disc profiles displayed in the figure are not simulation results but only illustrative sketches).
}. 
\label{schema}
\end{figure*}

We assume that the studied system has reached a steady state, both dynamically and collisionally. We consider the following forces: gravity and radiation pressure from the central star, gravitational pull from a perturbing planet and/or stellar companion. Collisions are assumed to produce bodies following a size distribution in $dN \propto s^{q}ds$. We divide particles into two categories: parent bodies (PB), i.e., bodies that are large enough not to be affected by radiation pressure, and small fragments steadily produced by collisions from these parent bodies and placed on eccentric or unbound orbits by radiation pressure. The magnitude of the radiation pressure force is quantified, as usual, by the parameter $\beta$, corresponding to the size-dependent ratio of this force to that of star's gravity. The dynamical evolution of the parent bodies and the smaller grains is computed using an adaptive-step 4$^{th}$ order Runge-Kutta integrator.

For the sake of simplicity, we consider a reference case of particles orbiting a central star and one external perturber on an eccentric orbit. Our numerical procedure is divided into three steps that can be schematically described as follows.

\subsection{Parent Body Runs} \label{pbr}

A first run is performed with only the parent bodies and thus only the central and perturbing gravitational forces. This run is stopped once a dynamical steady state has been reached, i.e., when no changes are observed between two consecutive passages of the perturber at the same phase. At this steady state, the position of the parent body disc is then recorded for a series of $n_{sav}$ different positions of the perturber on its orbit, separated by a fixed time interval $\Delta t_{sav}=t_{Porb}/n_{sav}$ (where $t_{Porb}$ is the orbital period of the perturber). \footnote{In the case, not studied here, of an interior perturber, this method might not sample the whole orbit of a parent body. But this is not a problem as long as the number of bodies is large enough to populate all of given body's orbit longitudes.} 

\subsection{Collisional runs} \label{cr}

A series of "collisional" runs are then performed, each taking as an initial condition the positions of the parent bodies for one of the $n_{sav}$ phases of the perturber, as indexed by the number $i_{\phi}$, with $1\leq i_{\phi} \leq n_{sav}$, from the sample of recorded PB configurations. For each collisional run, at $t=0$, $N_{num}$ particles are released following a size distribution in $dN \propto s^{q}ds$, with $q=-3.5$ \citep{dohn69}. The positions of these particles are then recorded, at regularly space time intervals $\Delta t_{sav}$ corresponding to the same sample of orbital phases of the perturber as in the PB run. At each integration timestep, each of the released particles is assigned a collisional destruction probability $f_{Dcoll}$ depending on its size $s$, velocity and the local geometrical optical depth $\tau_{r,\theta}$ of the system:
\begin{equation}
f_{Dcoll} = \left(\frac{s}{s_0}\right)^{\alpha} \frac{\delta V}{\delta V_0}\,\,\,\frac{4\pi\tau_{(r,\theta)}}{t_{orb}}\,dt,
\label{fcoll}
\end{equation}
where $dt$ is the time increment of the code, $t_{orb}$ the orbital period of a $\beta=0$ body at this location of the disc, $s_0$ is a reference size, $\delta V$ and $\delta V_0$ are the departures from the local Keplerian velocity $V_{Kep}$ of the considered particle and for a parent body with $\beta = 0$ respectively \citep[see][ for more details on the procedure]{theb10}. The optical depth values $\tau_{(r,\theta)}$ are derived from the $n_{sav}$ parent body runs, under the assumption that collisions occur mainly in the parent body region. This assumption is in agreement with previous debris disc modelling results \citep[e.g.][]{kriv06,theb07} and has been adopted in all studies investigating the outer regions of debris discs \citep[e.g][]{stru06,theb08}. In practice, we record the (steady-state) final positions of all parent body runs and transform them into $(r,\theta)$ maps of the vertical optical depth. These maps only give relative values and have to be normalized by $\left< \tau \right>$, the average normal optical depth assumed for the PB disc. The collisional activity can be thus tuned in by setting the initial value for $\left< \tau \right>$. Note that this procedure requires that $dt$ is smaller than the collision time for the grain to properly resolve the collisional destruction probability, a criteria that is always met for the study of perturbed debris discs, where $dt$ is a fraction of the orbital period and $t_{coll}$ is always $\geq$ $t_{orb}$.

The collisional runs are stopped when all particles have been eliminated, either by collisions or by dynamical ejection after an encounter with the perturber. At the end of this step, to each initial phase $i_{\phi}$ of the perturber is assigned a series of surface density maps $\sigma'(i_{\phi},0), \sigma'(i_{\phi},1),.., \sigma'(i_{\phi},j)$, recorded at the same regularly spaced time intervals $\Delta t_{sav}$, following the fate of all particles released when the perturber was at the $i_{\phi}$ position. Note that all snapshots $\sigma'(i_{\phi},j)$, $\sigma'(i_{\phi},j+n_{sav})$, $\sigma'(i_{\phi},j+2n_{sav})$, etc., correspond to successive passages of the perturber at the same orbital phase (each perturber orbit being divided into $n_{sav}$ positions).

\subsection{Recombining} \label{rec}

In the final stage, we use all data collected in step 2 to reconstruct the
dust distribution, \emph{at steady state}, for each possible orbital phase of the perturber. The principle is that, at a given time $t_{i_{\phi}}$ corresponding to the perturber's position $i_{\phi}$, the total dust population regroups grains that have just been produced (for the present position $i_{\phi}$ of the perturber), as well as survivor grains that have been produced earlier (when the perturber was at a different orbital position), and that have not been collisionally destroyed or dynamically ejected yet. The procedure is the following: we start with the most recent dust particles, produced at $t_{i_{\phi}}$, whose spatial distribution is given by the saved record $\sigma'(i_{\phi},0)$. We then add the previous generation, produced at $t_{i_{\phi}}-\Delta t_{sav}$ when the perturber was at angular position index $i_{\phi}-1$, whose spatial distribution at time $t_{i_{\phi}}$ (i.e., at time $\Delta t_{sav}$ after their release) is given by the saved record $\sigma'(i_{\phi}-1,1)$. This procedure is then iterated, working our way back in time and piling up all the surviving grains from the successive records $\sigma'(i_{\phi}-j,j)$, to produce the total geometrical optical depth at time $t_{i_{\phi}}$
\begin{equation}
 \sigma(i_{\phi}) = \sum_{j=0}^{j=j_{max}} \sigma'(i_{\phi}-j,j)
\end{equation}

The index $j_{max}$ corresponds to the most ancient record, for which all initially released particles have been eliminated after the time $j_{max}\times \Delta t_{sav}$ separating their release from the present time. Note that all the snapshots $\sigma'(i_{\phi}-j,j)$, $\sigma'(i_{\phi}-j-n_{sav},j+n_{sav})$, $\sigma'(i_{\phi}-j-2n_{sav},j+2n_{sav})$, etc., are taken from the same source collisional run, with the perturber position at release being $(i_{\phi}-j)$, but separated by an integer number of orbital periods. In other words, the $\sigma'(i_{\phi}-j-k.n_{sav},j+k.n_{sav})$ record corresponds in practice to the $\sigma'(i_{\phi}-j,j+k.n_{sav})$ one.

An illustrated summary of the main steps of our numerical procedure can be found in Fig.\ref{schema}.

\subsection{Tests}

\begin{figure*}
\makebox[\textwidth]{
\includegraphics[scale=0.5]{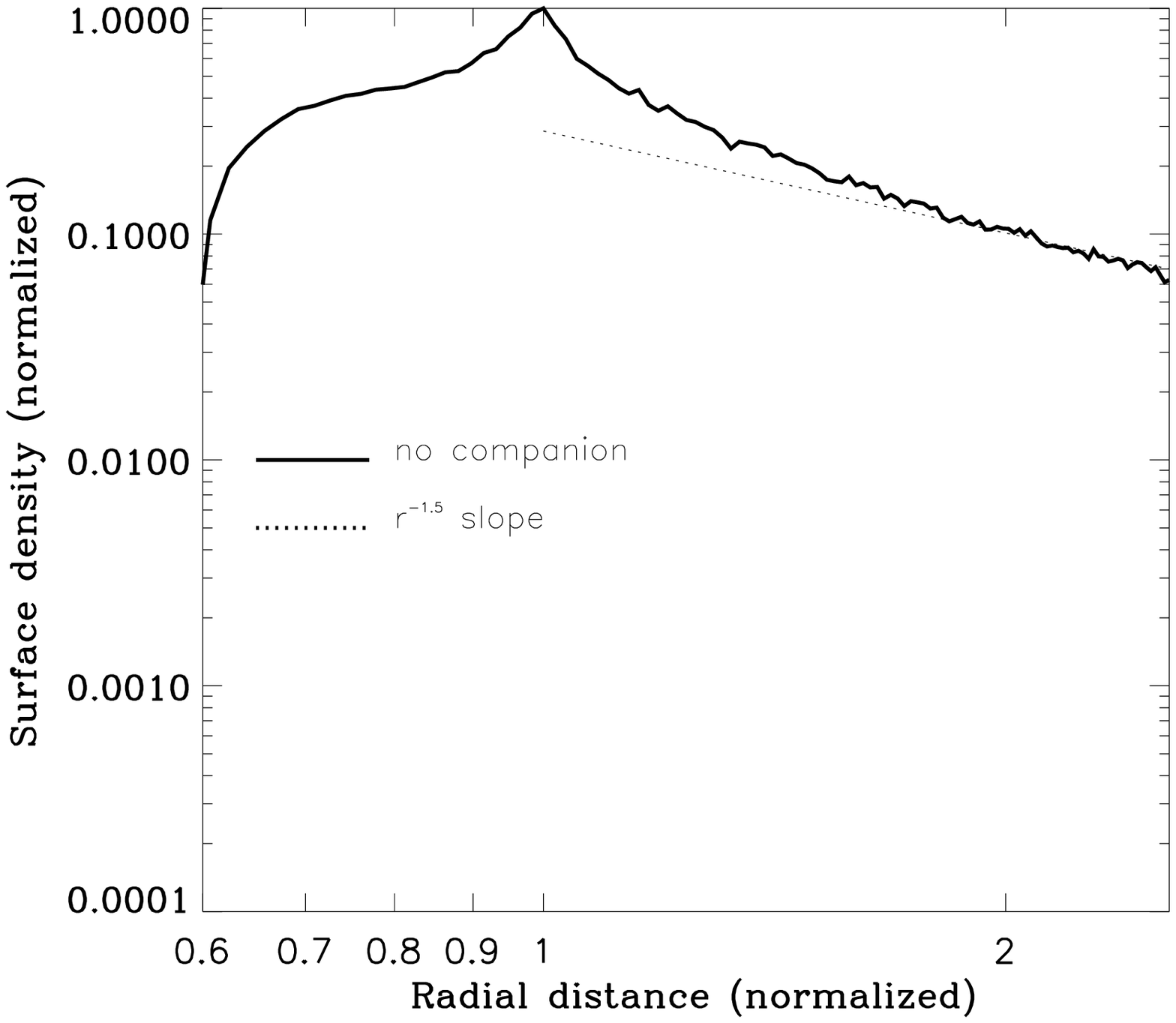}
\includegraphics[scale=0.5]{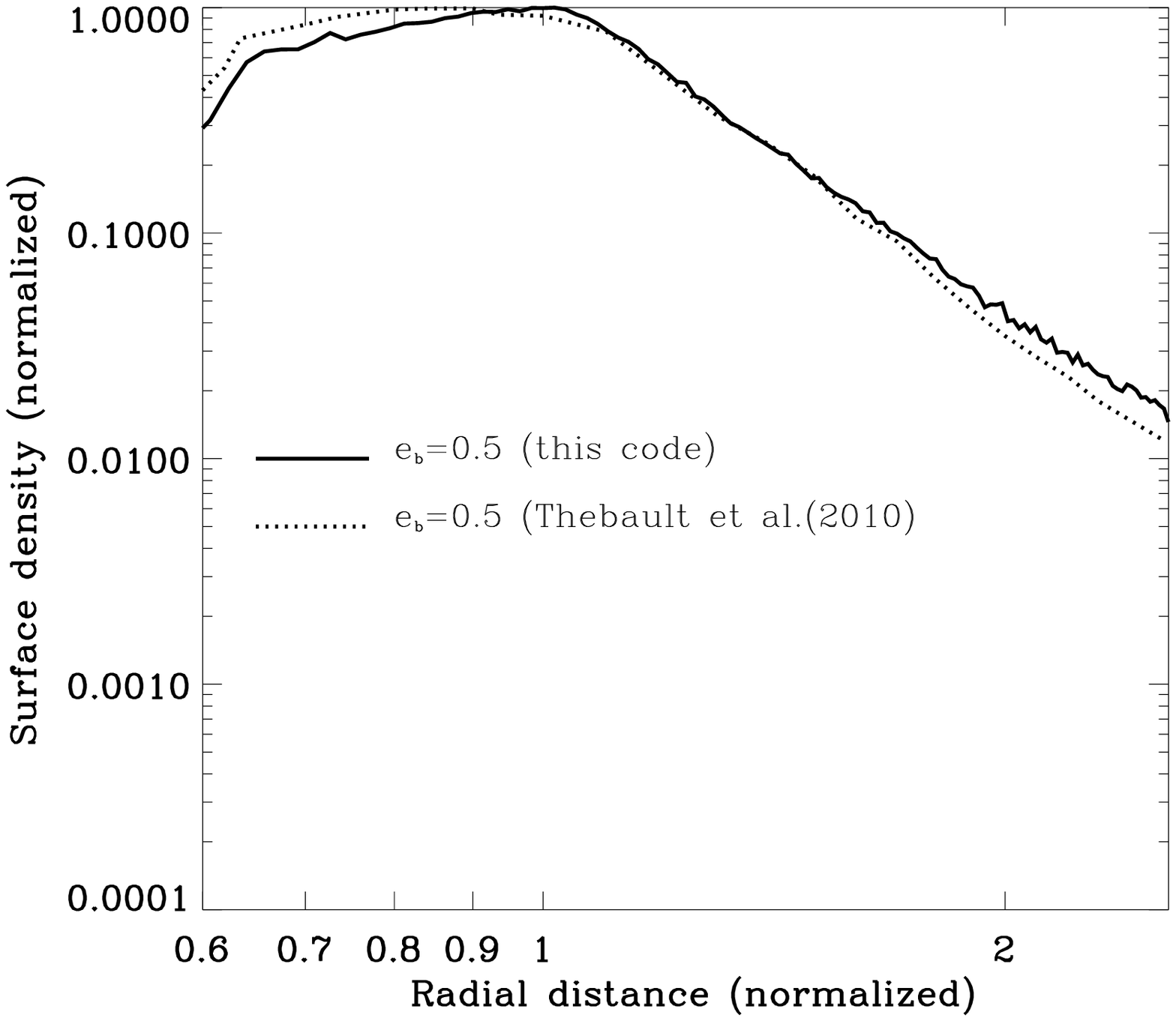}
}
\caption[]{ Test runs: \emph{Left Panel}: Azimutally averaged radial profile of the normalized surface density for a fiducial case with no companion. The expected theoretical asymptotic behaviour in $r^{-1.5}$ is plotted as a reference. The initial extension of the parent body ring is set to $r=1$
\emph{Right Panel}: Azimutally averaged radial profile of the normalized surface density for a companion of eccentricity $e_b=0.5$ as compared to the results of \citet{theb10} with a different code. For this with-companion run, disntances have been normalized to the theoretical limit for orbital stability $r_{crit}$ (see Sec.\ref{setup}).

}
\label{test}
\end{figure*}

As there is no analytical solution to the full problem of a collisionally active disc perturbed by a massive body, we follow an approach similar to that of \citet{star09} and first check the correctness of our algoritm's solution for a simplified case with no companion. For this case, \citet{stru06} and \citet{theb08} have shown that, for a collisional debris ring whose parent bodies distribution has a sharp outer edge $r_{out}$, the radial profile of the surface density beyond $r_{out}$ should tend towards an aymptotic slope in $r^{-1.5}$ (because of the small grains produced within the ring and pushed there on eccentric orbits). Note that we do not try to reproduce the numerical simulations of \citet{stru06} or \citet{theb08}, which did address slighly different issues, but the robust conclusions from their analytical derivations showing that a collisional ring always tends towards a profile in -1.5 outside the collisional active region. We thus run a test simulation with no companion and a parent body ring having a sharp outer edge at $r_{out}=1$ (arbitrary unit). As is clearly seen in Fig.\ref{test}a, we do obtain a profile that tends towards the standard slope in -1.5 beyond $r\sim 1.5$.  
Note that this test is not a validation of the whole collisional procedure, as the -1.5 slope is a result that does only weakly depend on the collisional rate within the PB ring. However, it validates the recombination procedure, which is rather complex, piling up maps after maps coming from different runs and at different times, none of these individual maps having a radial profile tending towards -1.5.

For the case with a massive exterior companion, we perform an additional test by comparing our results to the ones obtained by \citet{theb10} who, using a completely different algortim with only 1-D resolution, estimated the azimutally averaged radial profile of the surface density. We take as a reference the case with a $M_2=0.25 M_1$ companion on a $e_B=0.5$ eccentric orbit and a parent body ring of optical depth $\tau=2\times 10^{-3}$  extending out to the orbital stability limit $r_{crit}$ (see Sec.\ref{setup}). We then azimutically average the spatial distribution and display it in Fig.\ref{test}. As can be seen in Fig.\ref{test}b the obtained averaged radial profile is a very good match to the one obtained by \citet{theb10} with their different method (see Fig.2 of that paper), i.e. a steeper profile than in the previous no-perturber case because of the steady removal of high-$\beta$ particles by the companion. This second test is a robust validation of the collisional and dynamical prescriptions, as the shape of the radial profile directly depends on the balance between the collisional activity in the PB disc and the dynamical ejection of particles perturbed by the companion.

\section{A case study: debris disc in a binary} \label{res}

To illustrate the potential of our code, we consider a case study of a circumprimary debris disc in a binary system, of particular interest in light of the recent surge of research regarding planet formation in binaries \citep[for a recent review, see][]{theb11}. This case has been investigated in several recent works, both observational \citep{tril07,plav09,duch10} and theoretical \citep{theb10}. The main issue investigated in these studies was the extent of circumprimary discs, in particular if the companion star can induce a truncation that can be detectable when looking at infra-red excesses or at the radial profile of the resolved disc. 
The numerical study of \citet{theb10} showed that the coupling of collisional activity and radiation pressure plays a crucial role, steadily placing small dust grains in regions that are in principle dynamically unstable. Since these grains dominate the flux at all wavelengths up to mid-IR, debris discs can thus appear to extend far beyond the theoretical radial distance $r_{crit}$ for orbital stability around the primary.
However, \citet{theb10} used a collisional code with only 1-D spatial resolution (all azimutal information being lost in phase averaging), and were thus unable to study how binarity affects the \emph{shape} of circumstellar discs. This issue is crucial, as the presence of a companion star has been invoked as a possible explanation for several systems' aspect, such as HR4796 \citep{auge99,schn09} or HD141569 \citep{auge04,quil05,ardi05}.

\subsection{Setup} \label{setup}

\begin{table}
\begin{minipage}{\columnwidth}
\caption[]{Set up for the disc-in-a-binary runs}
\renewcommand{\footnoterule}{}
\label{init}
\begin{tabular*}{\columnwidth} {ll}
\hline
PARENT BODY RUN &  \\
\,\,\,Number of test particles & $ N_{PB}=2 \times 10^{5}$\\
\,\,\,Initial radial extent \footnote{normalized to $r_{crit}$} & $0.6<r<1.1$\\
\,\,\,Initial eccentricity & $0.01\leq e_0 \leq 0.05$\\
COLLISIONAL RUNS & \\

\,\,\,Average optical depth \footnote{to derive $f_{Dcoll}$ (Equ.\ref{fcoll})} & $\left< \tau \right>= 2\times10^{-3}$\\
\,\,\,Number of test particles & $ N_{num}=2 \times 10^{5}$\\
\,\,\,Size range \footnote{as parameterized by $\beta \propto 1/s$} & $\beta(s_{max})=0.012\leq\beta(s)\leq\beta(s_{min})=2.5$\\
\,\,\,Size distribution at $t=0$ & $dN(s) \propto s^{-3.5}ds$\\
\hline
\end{tabular*}
\end{minipage}
\end{table}

We consider a binary of separation $a_B$, eccentricity $e_B$ and mass ratio $\mu$. We make the not-so-unreasonable assumption that the disc of large \emph{parent bodies} has been shaped and truncated by the companion star. To ensure this, we allow the initial disc to extend slightly beyond the empirical (1-D) outer limit $r_{crit}$ for orbital stability derived by \citet{holm99}, and let the code naturally remove all unstable particles. 

To allow easy comparison between different cases, all distances are normalized so that $r_{crit}=1$. In order to avoid the huge computational cost of calculating orbits very close to the primary, and since these regions are in any case the least affected by binarity, we consider a ring-like configuration for the initial disc: $0.6\leq r \leq 1.1$ (in units of $r_{crit}$).

For the binary, we consider a fixed mass ratio $\mu=0.25$, corresponding to estimates of the mean mass ratio in binaries derived from extensive surveys \citep{kroupa90,duq91}, and explore 4 different orbital configurations: $e_B = 0$, $e_B = 0.2$, $e_B = 0.5$ and $e_B = 0.75$ (the values of $a_B$ are then automatically obtained from the requirement that $r_{crit}=1$).

We consider an average vertical optical depth $\left< \tau \right>=2\times10^{-3}$, typical of dense debris discs like $\beta$ Pictoris.

For the parent body runs, all particles start on circular orbits ($e=0$) in the binary's midplane ($i=0$). For the collisional runs, particles are released from the parent bodies seeds following a size distribution in $dN(s) \propto s^{-3.5}ds$. Particle sizes are parameterized by the value of their $\beta$ parameter. Sizes are distributed continuously between $\beta_{max}=0.5$ and $\beta_{min}=0.012$. We take the minimum particle size to be $\beta_{max}=0.5$, corresponding to the smallest grains on bound orbits. The contribution of smaller grains, below the blow-out size limit, is negligible for systems with optical depths in the $\left< \tau \right> \sim 10^{-3}$ range \citep[see Fig.3 of][]{theb10}.
The number of sampled positions within a binary orbit is $n_{sav}=10$. Convergence test runs have shown that this is the optimal value: higher values do not yield significant changes in the final recombined density maps, while runs with lower $n_{sav}$ lead to diverging results. Note that for the first binary orbit of the collisional run, the sample value is higher, $n_{sav}=100$, in order to accurately track the rapid initial blow-out of high-$\beta$ grains.
All initial parameters are summarized in Tab.\ref{init}.

\subsection{Results} \label{resu}

\begin{figure*}

\makebox[\textwidth]{
\includegraphics[scale=0.365]{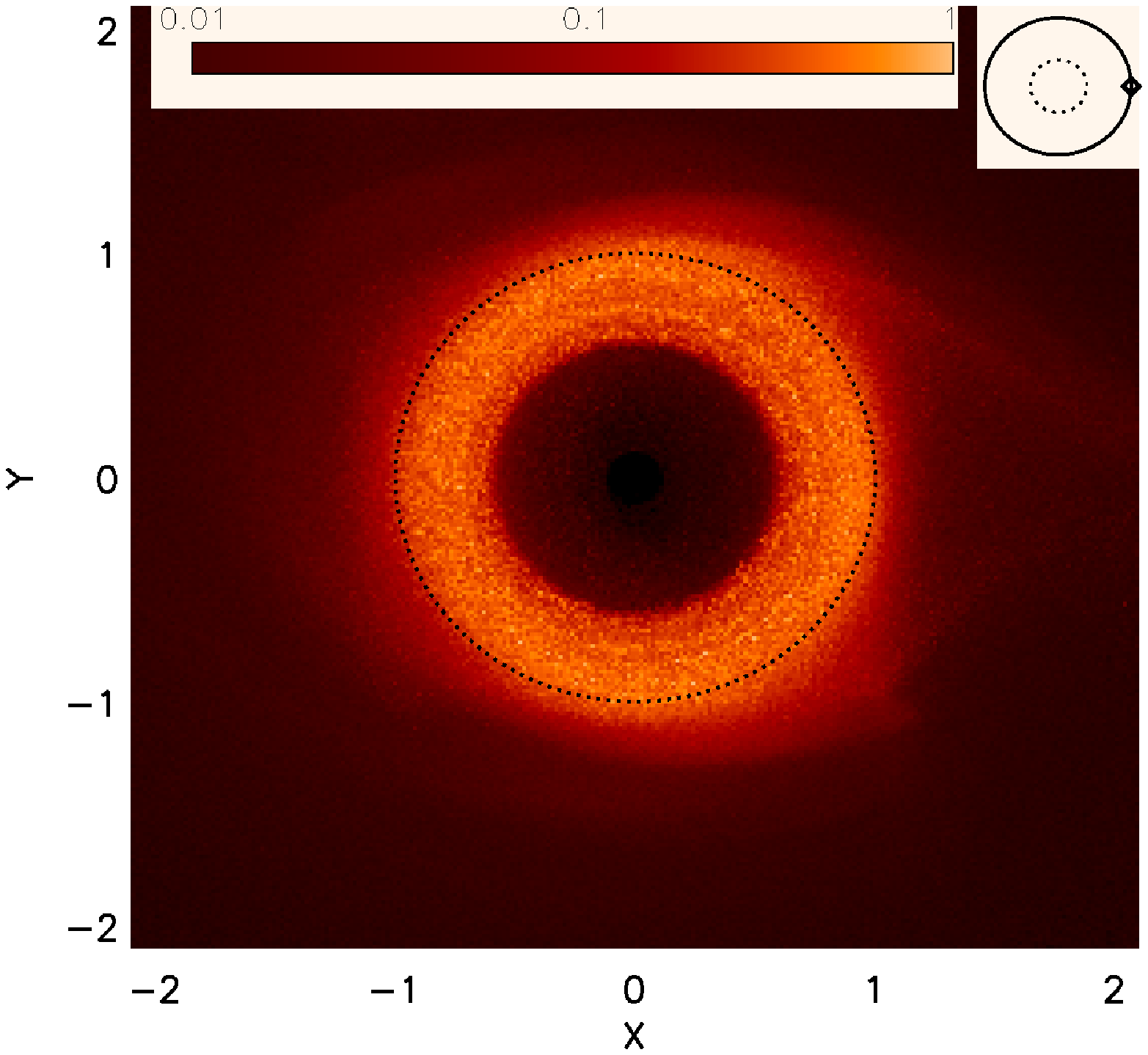}
\includegraphics[scale=0.365]{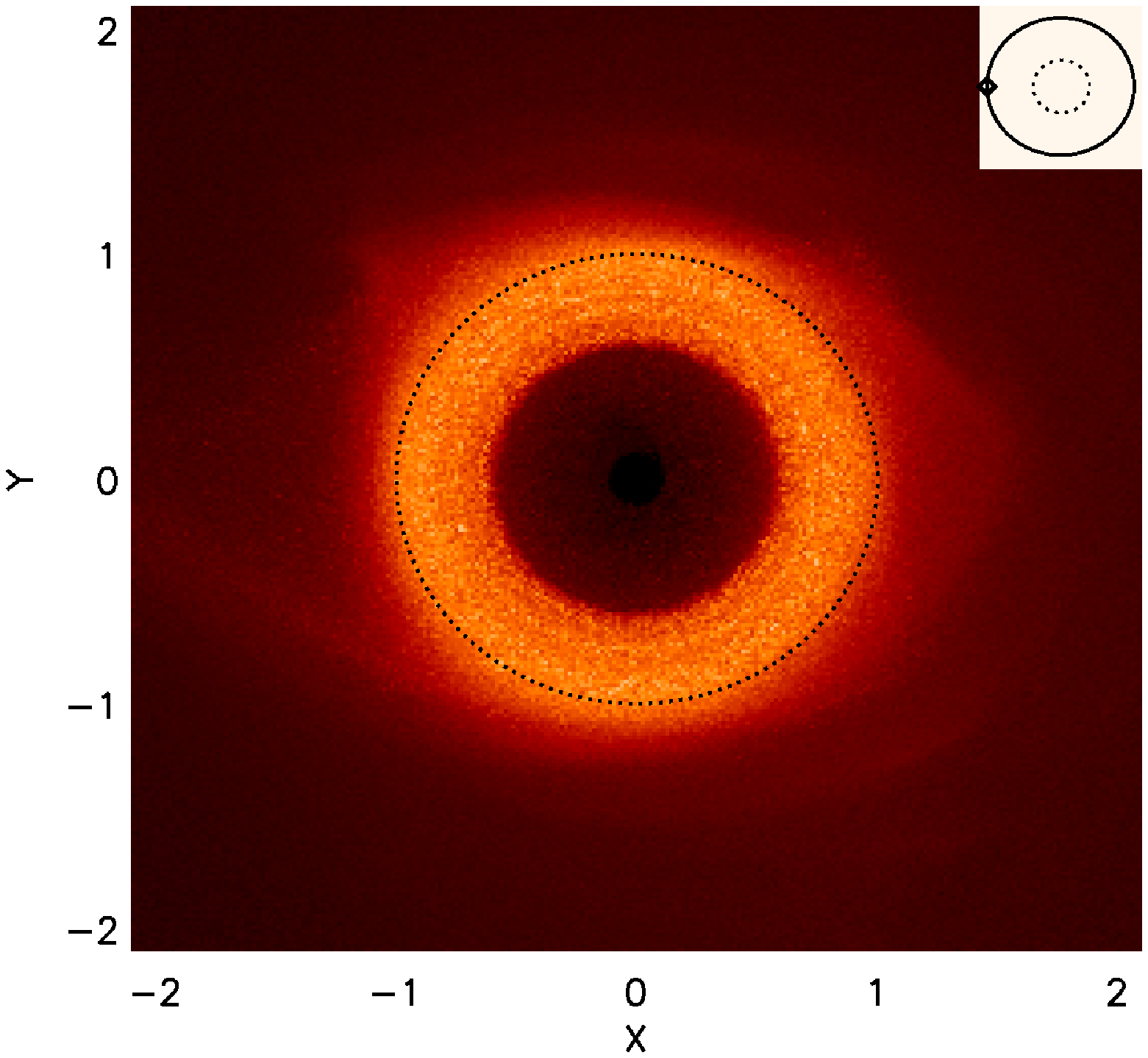}
}

\makebox[\textwidth]{
\includegraphics[scale=0.365]{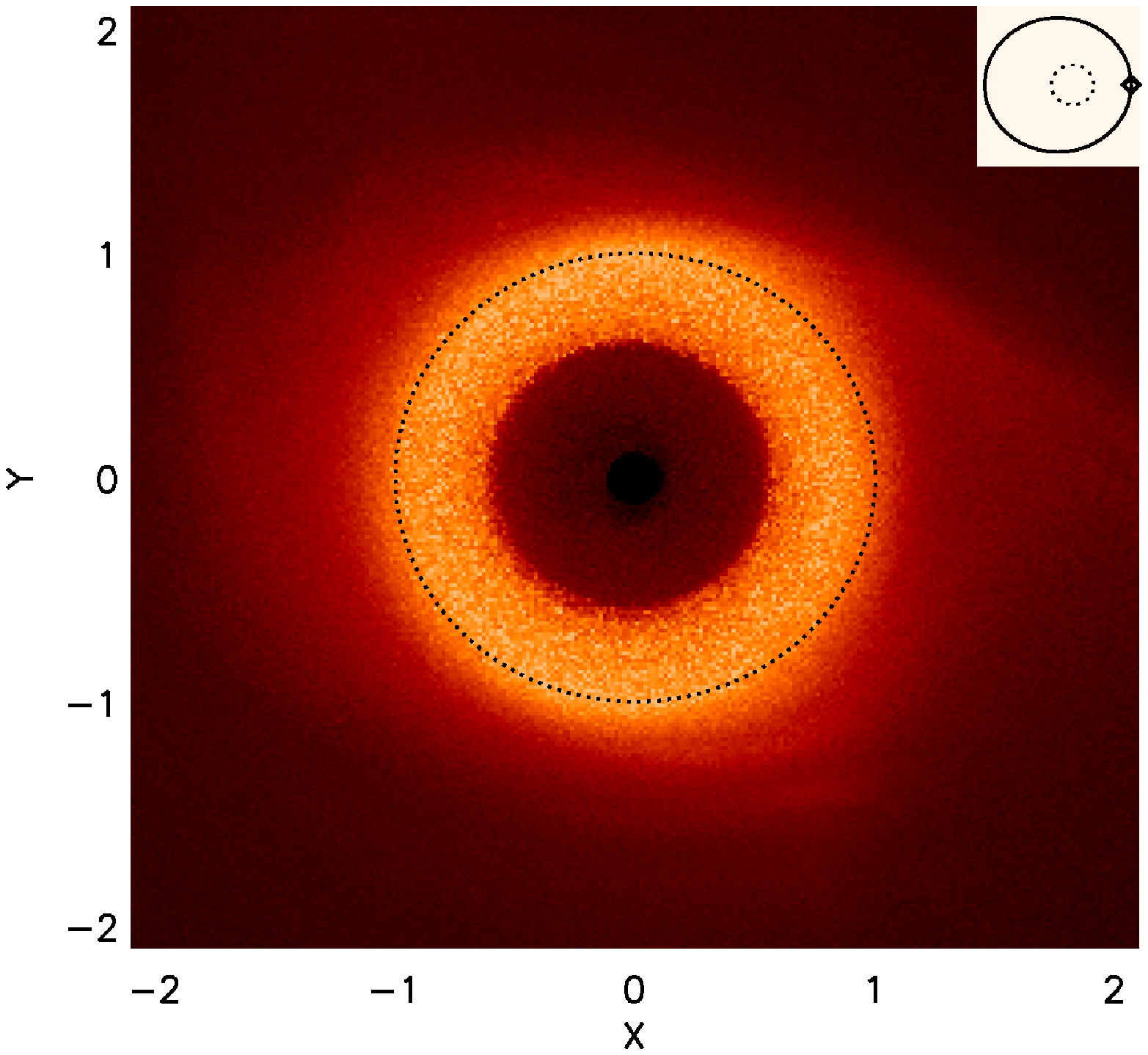}
\includegraphics[scale=0.365]{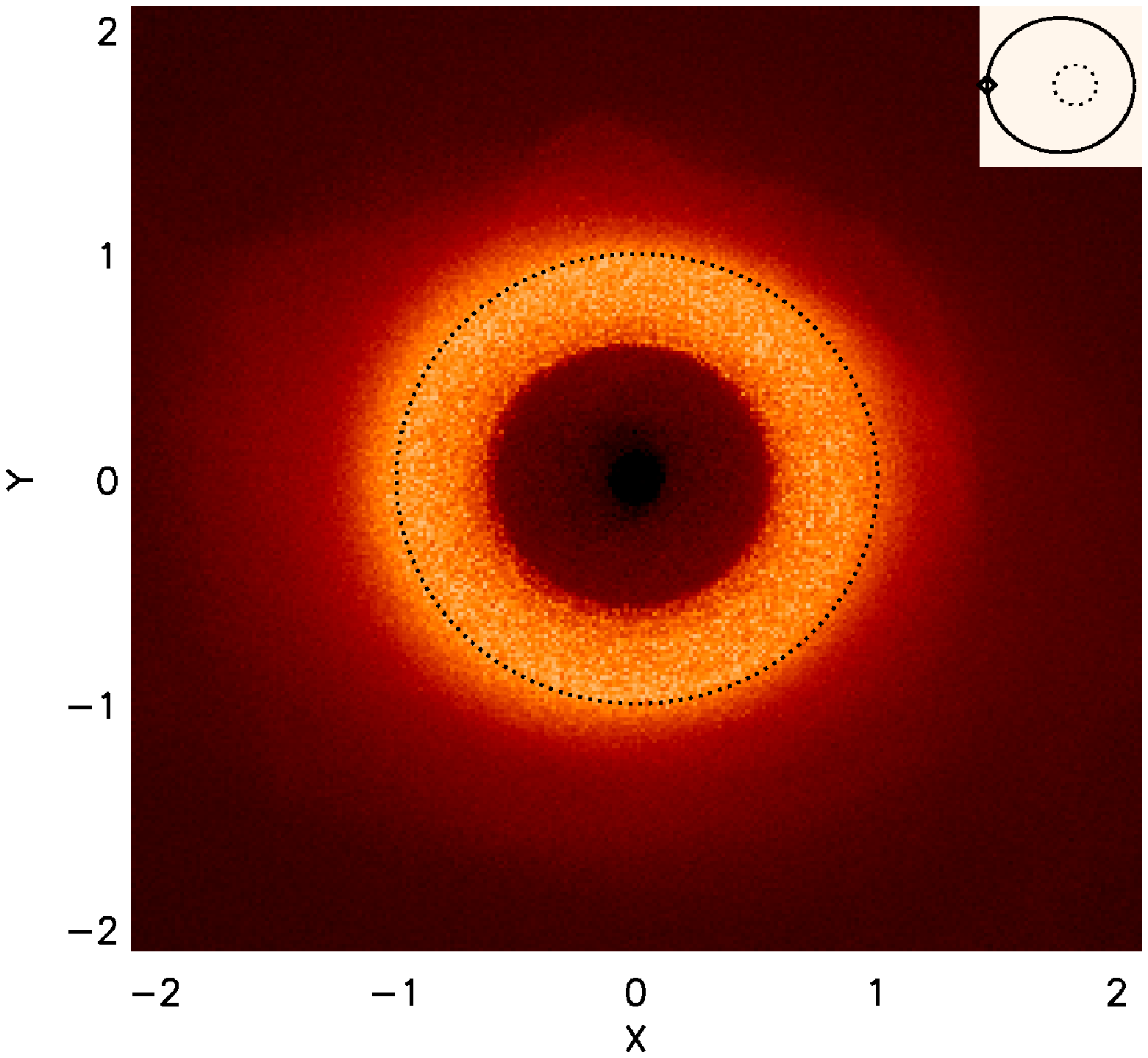}
}

\makebox[\textwidth]{
\includegraphics[scale=0.365]{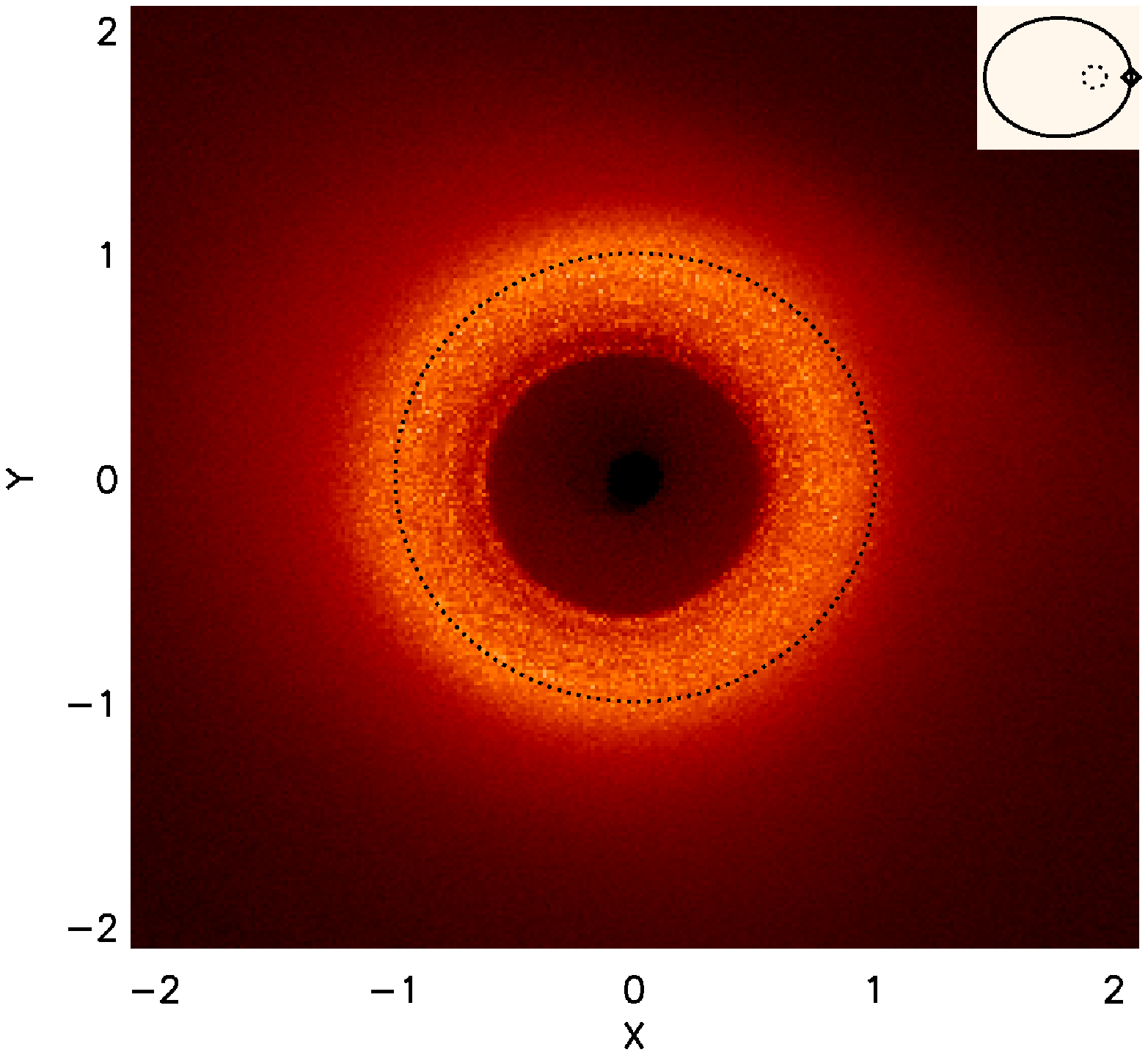}
\includegraphics[scale=0.365]{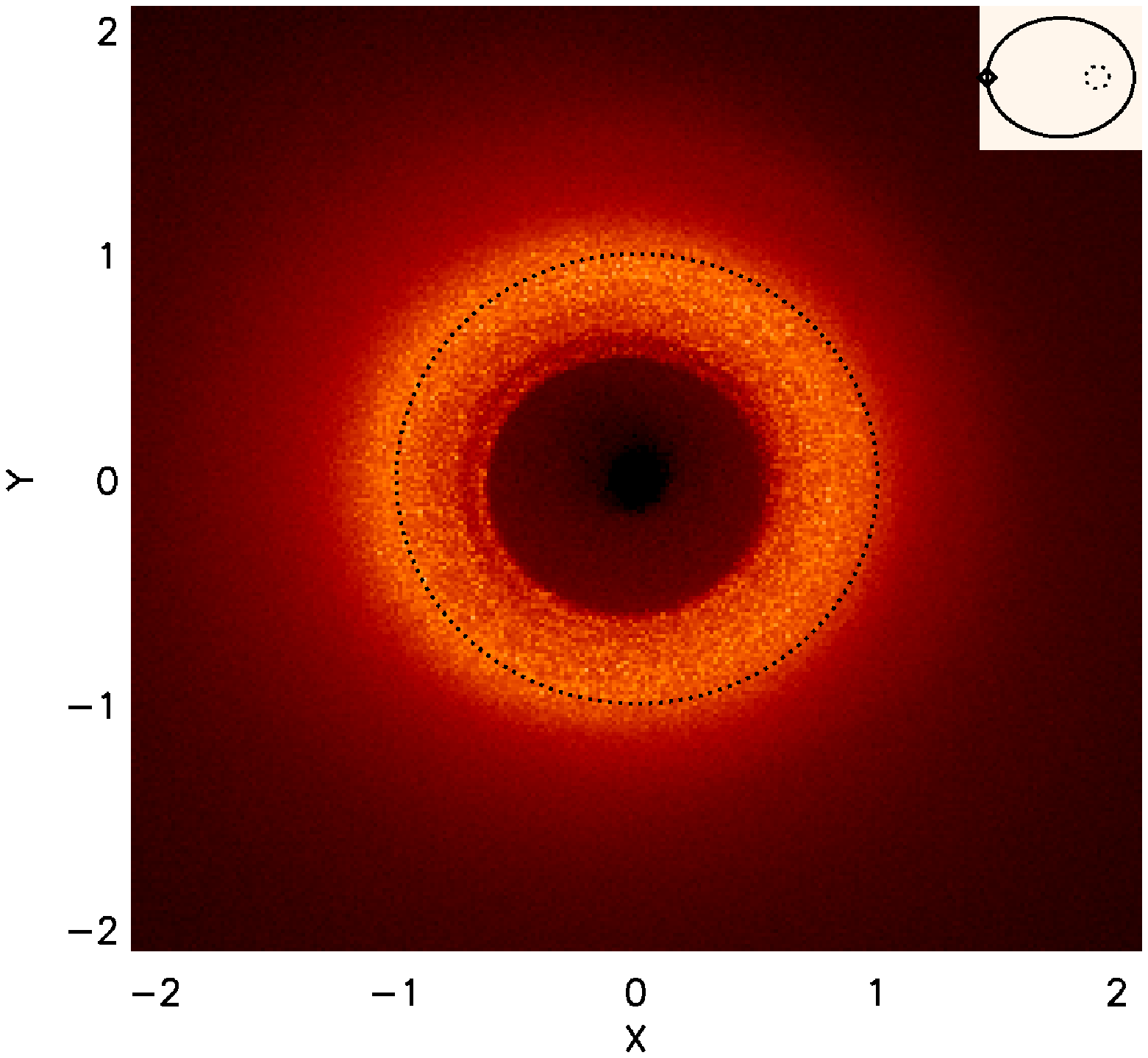}
}

\makebox[\textwidth]{
\includegraphics[scale=0.365]{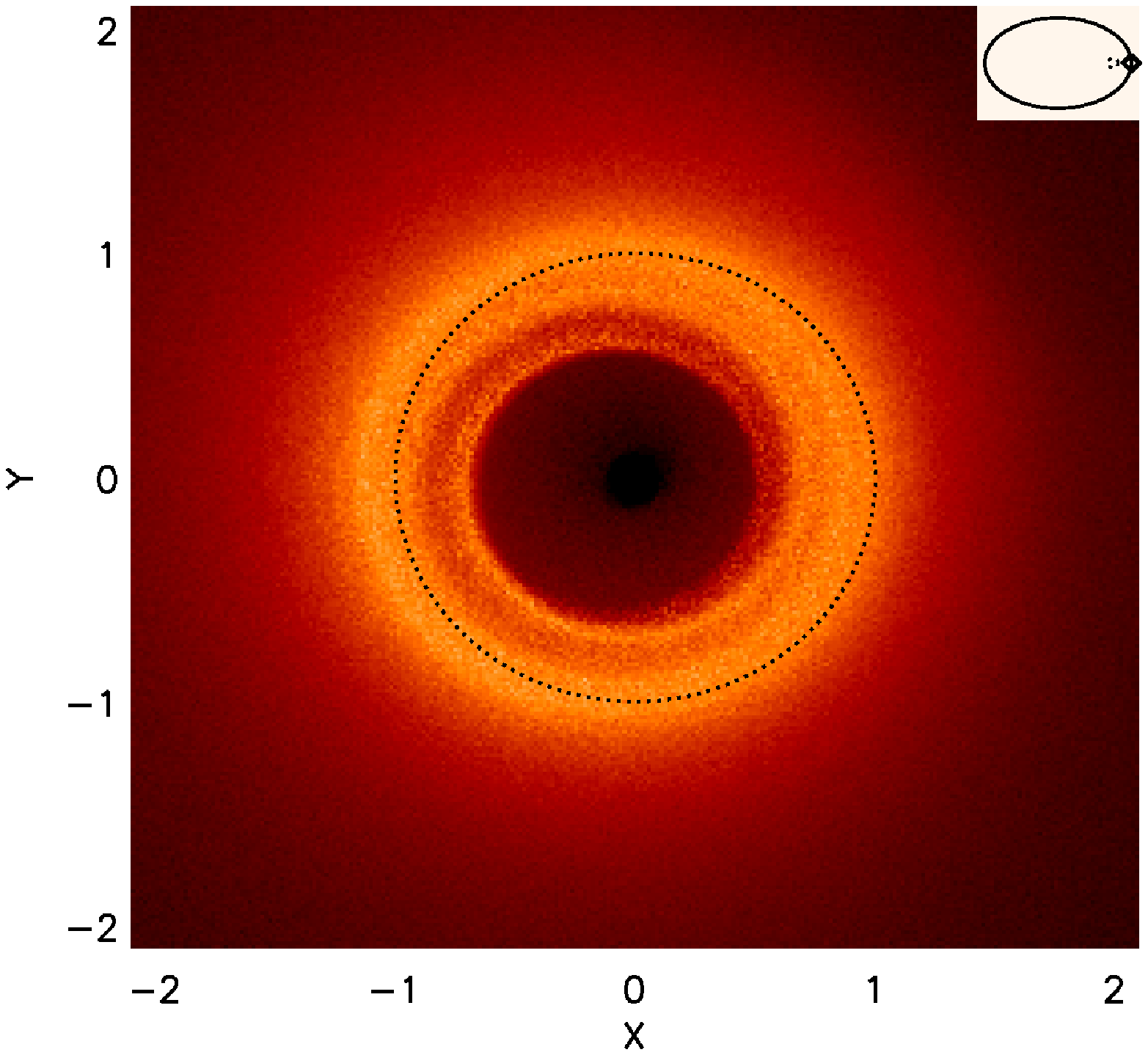}
\includegraphics[scale=0.365]{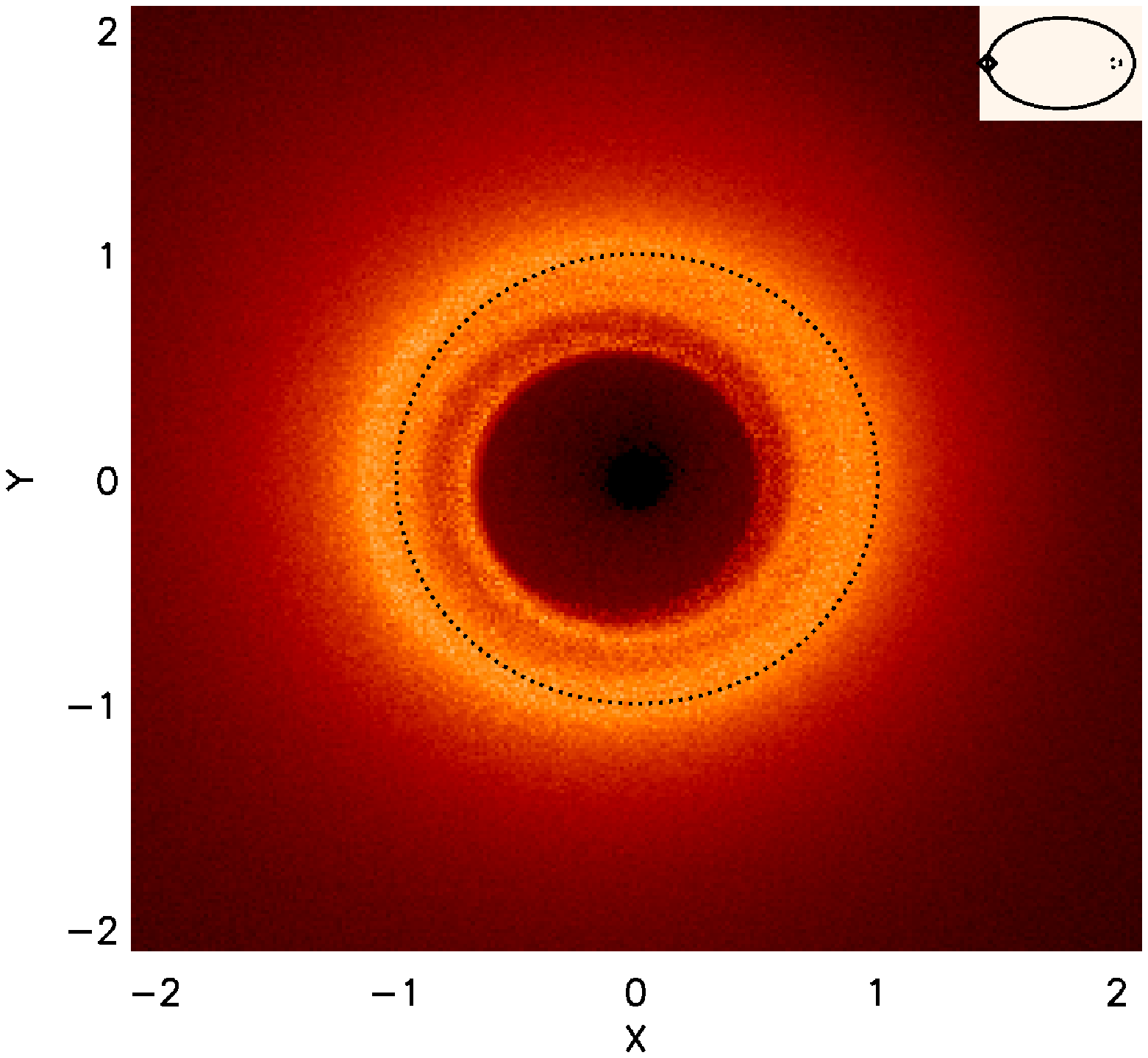}
}
\caption[]{Normalized vertical optical depth, at collisional and dynamical steady state, for  4 binary configurations; from top to bottom: $e_B=0$, $e_B=0.2$, $e_B=0.5$ and $e_B=0.75$. For each case, the disc profile is shown at periastron (left) and apoastron (right) passages of the companion star. The dotted circle is the theoretical outer radius for orbital stability $r_{crit}$. The rectangle in the upper right corner shows the orbit of the binary as compared to the $r_{crit}$ radius (dotted circle) and the current position of the companion on its orbit. (An animated version of these results, displaying disc profiles for 10 different orbital positions of the companion, can be found at http://lesia.obspm.fr/perso/philippe-thebault/bindeb.html).
}
\label{final}
\end{figure*}

Depending on the considered configuration, the parent body simulations are run for 200 ($e_B=0$ case) to 500 ($e_B=0.75$) orbital periods of the binary in order to reach a steady state. This steady state is reached once all particles on unstable orbits have been removed and once all transient dynamical features have disappeared. As shown by \citet{auge04}, the most notable of these transient features are spiral arms, which develop because of the sudden introduction of a perturbing body, and disappear on the scale of a few hundred orbital periods.
We then perform the collisional runs following the procedure presented in Sec.\ref{cr}. The final disc profiles, after recombination of all collisional runs following the method described in Sec.\ref{rec}, are displayed in Fig.\ref{final}.

A first important point is that these results confirm the main conclusion of \citet{theb10} that circumprimary debris discs extend far outside the outer limit for orbital stability. For all four binary configurations the regions beyond $r_{crit}=1$ are populated by small grains steadily produced by collisions in the parent body ring and placed on eccentric orbits by radiation pressure. The level of dustiness in these "forbidden" regions increases with $e_B$ because, for highly eccentric binaries, the perturbing companion is far from the main ring between two passages at periastron, so that small grains placed by radiation pressure in the $r\geq r_{crit}$ region have a longer dynamical survival timescale before being removed.

\begin{figure*}
\makebox[\textwidth]{
\includegraphics[scale=0.365]{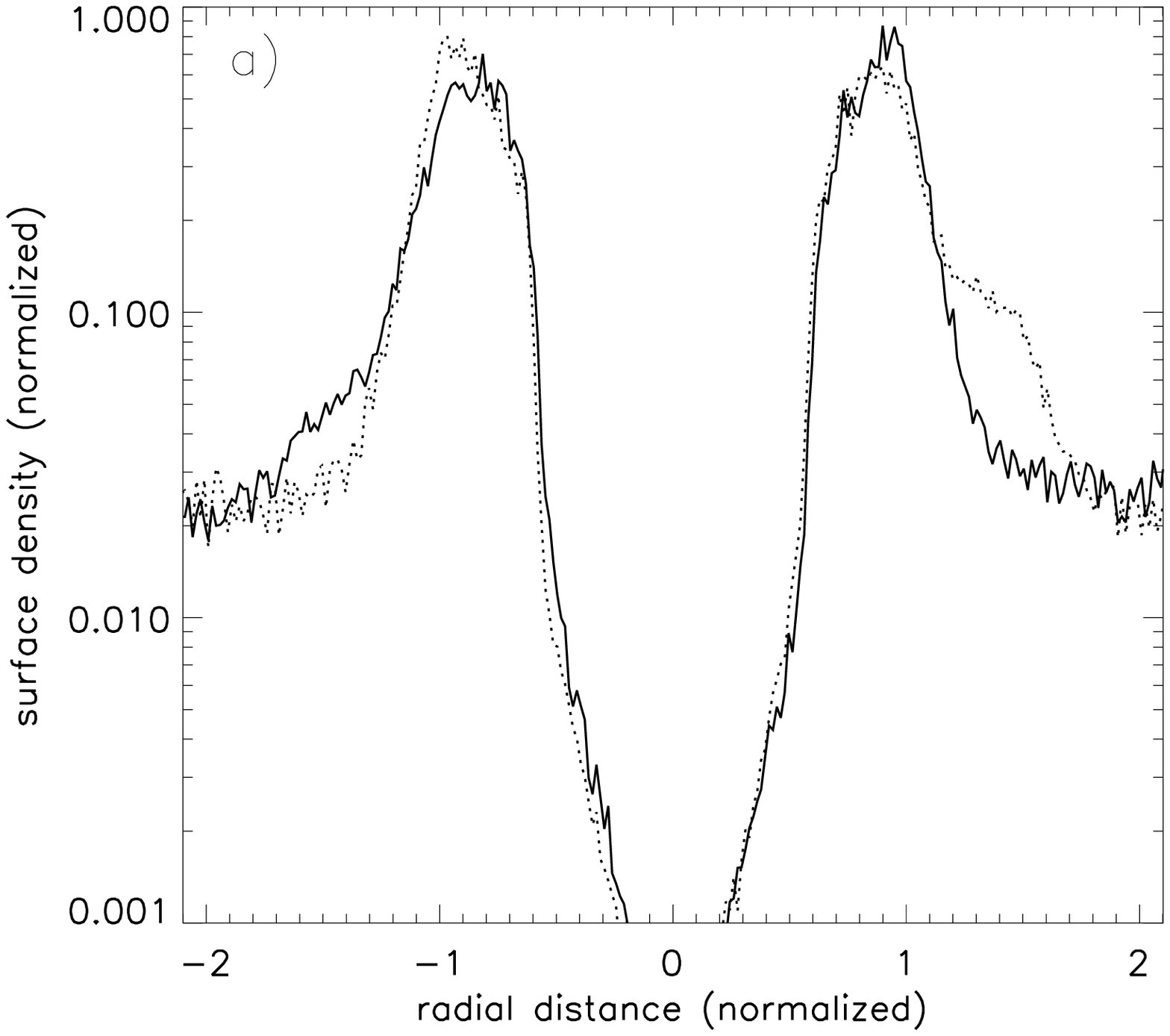}
\includegraphics[scale=0.365]{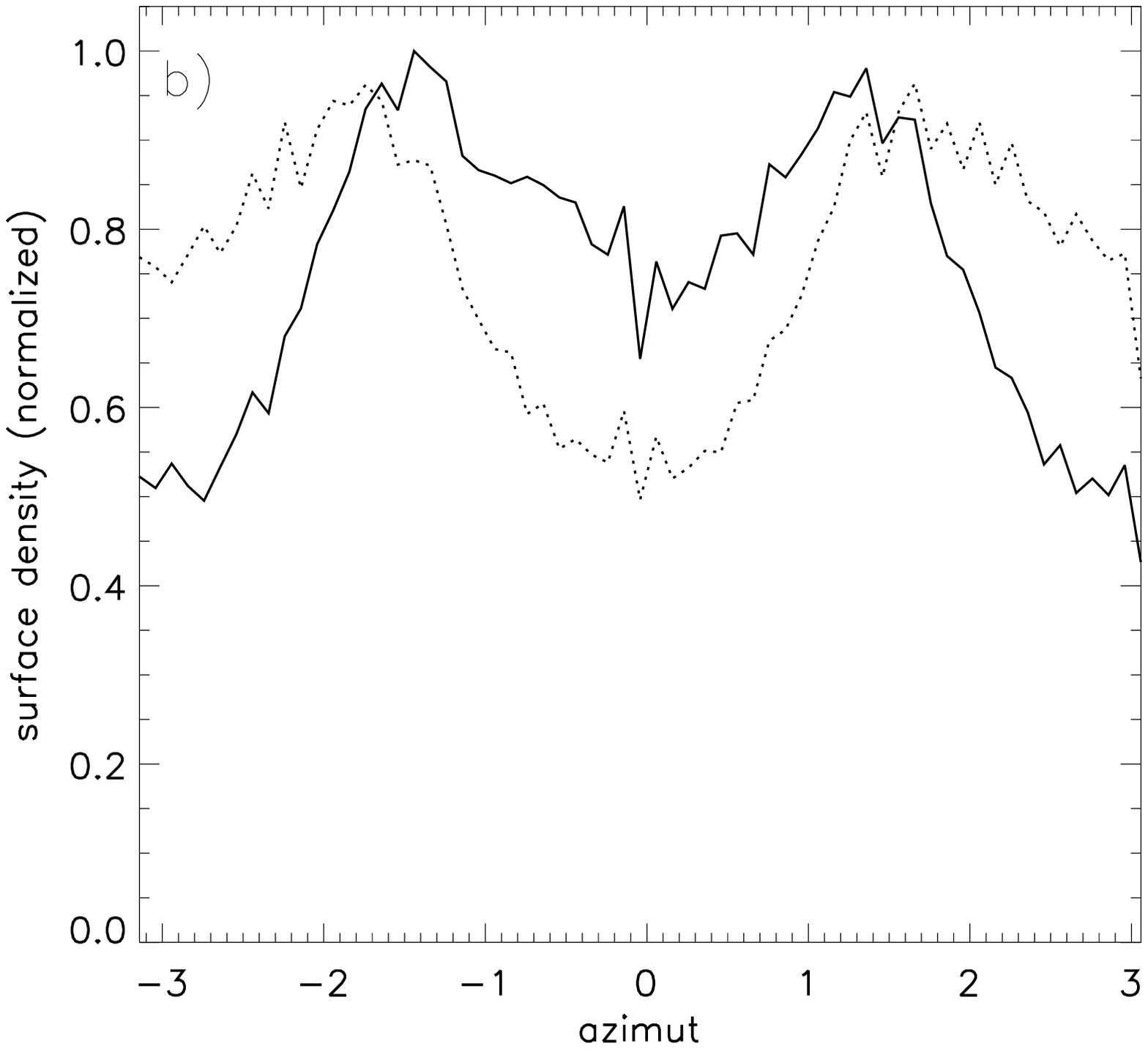}
}
\makebox[\textwidth]{
\includegraphics[scale=0.365]{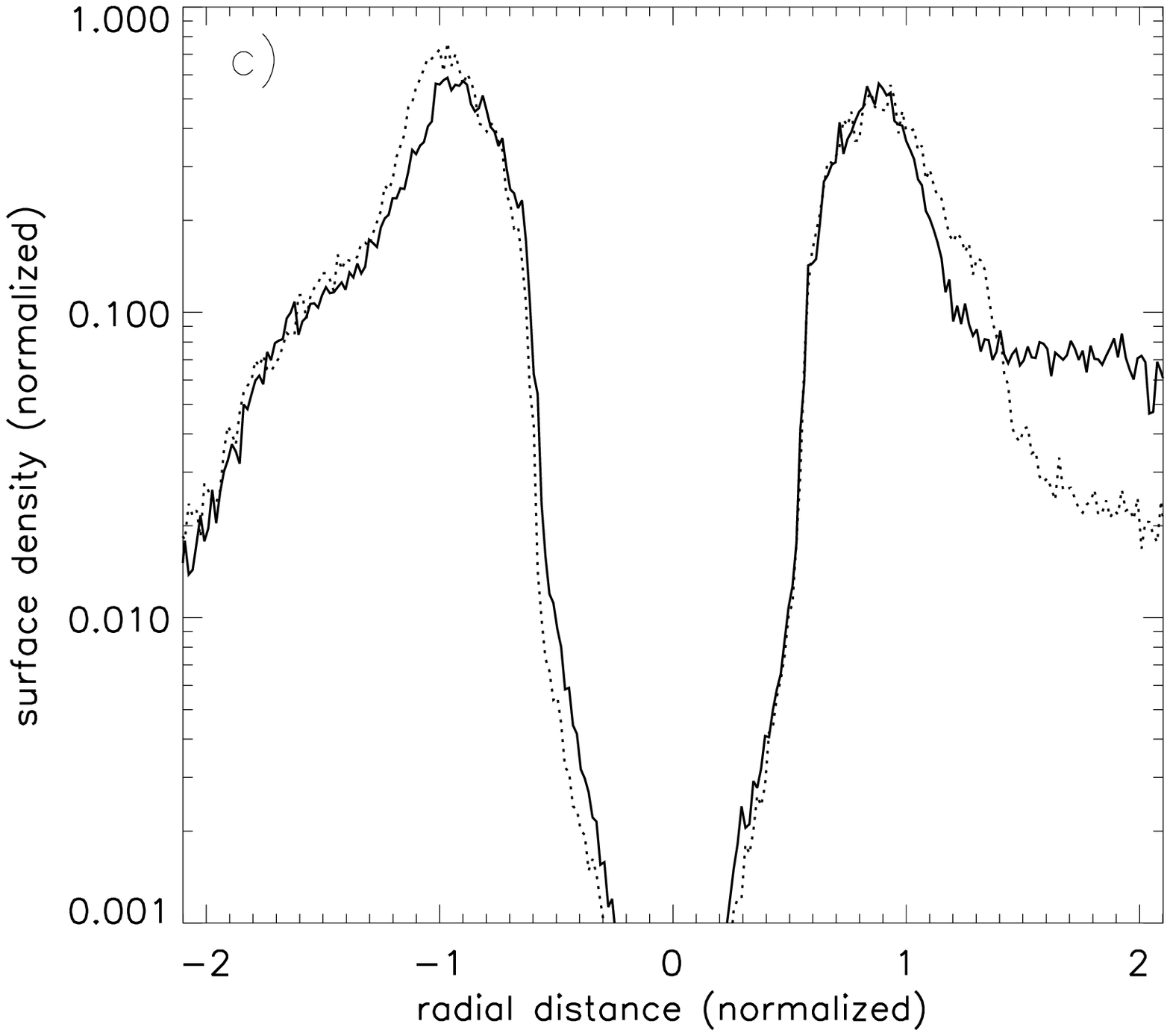}
\includegraphics[scale=0.365]{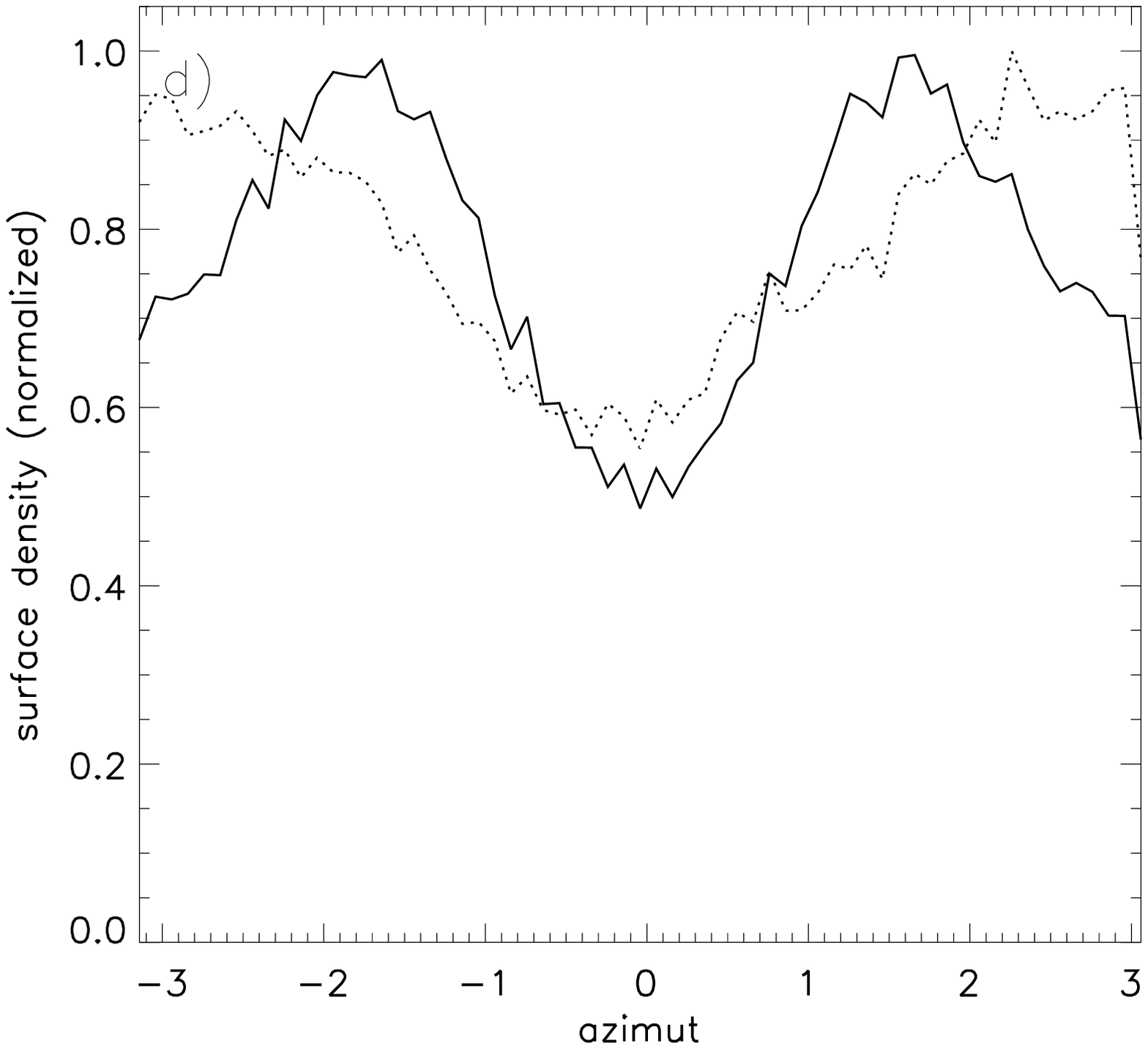}
}
\makebox[\textwidth]{
\includegraphics[scale=0.365]{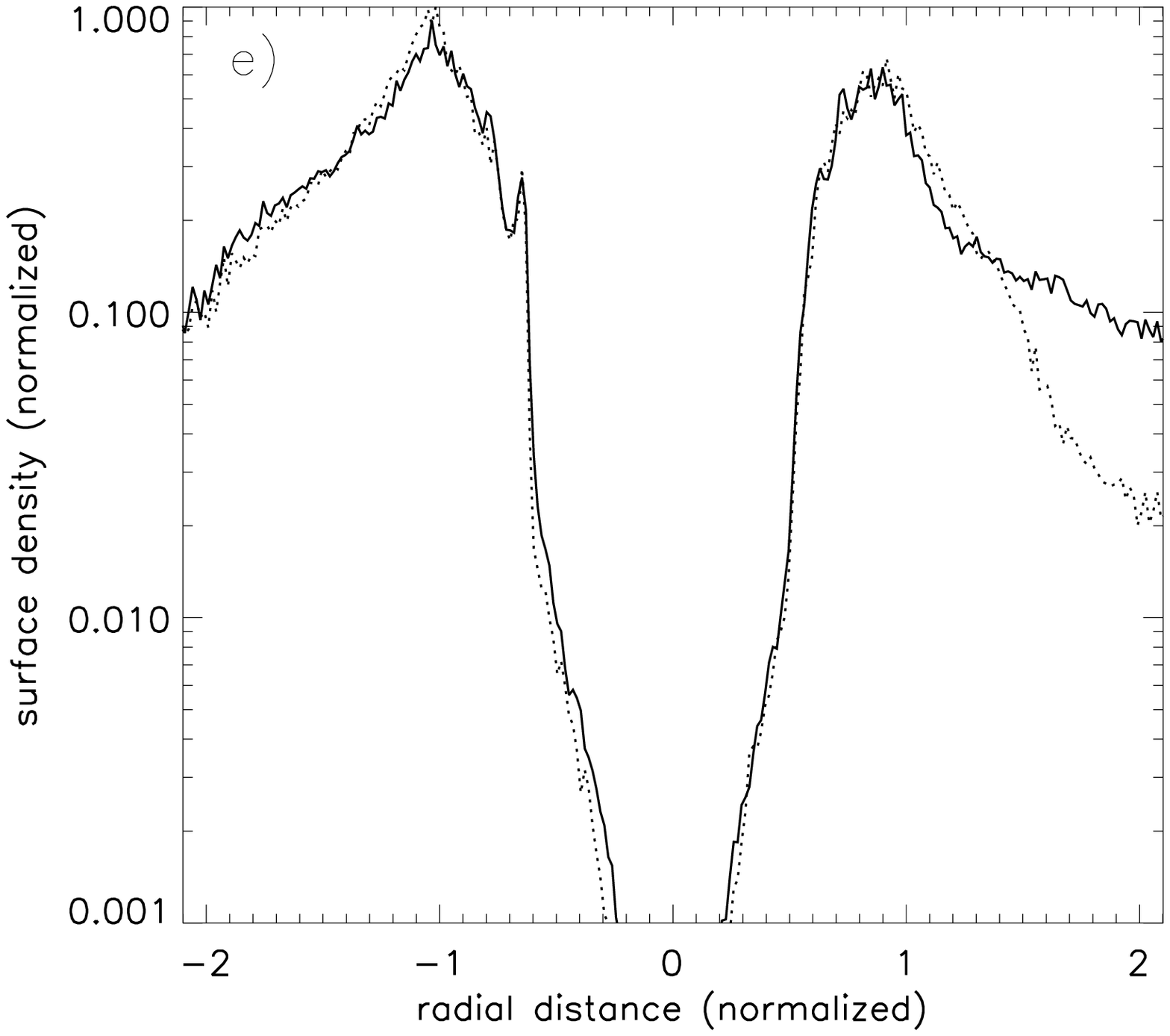}
\includegraphics[scale=0.365]{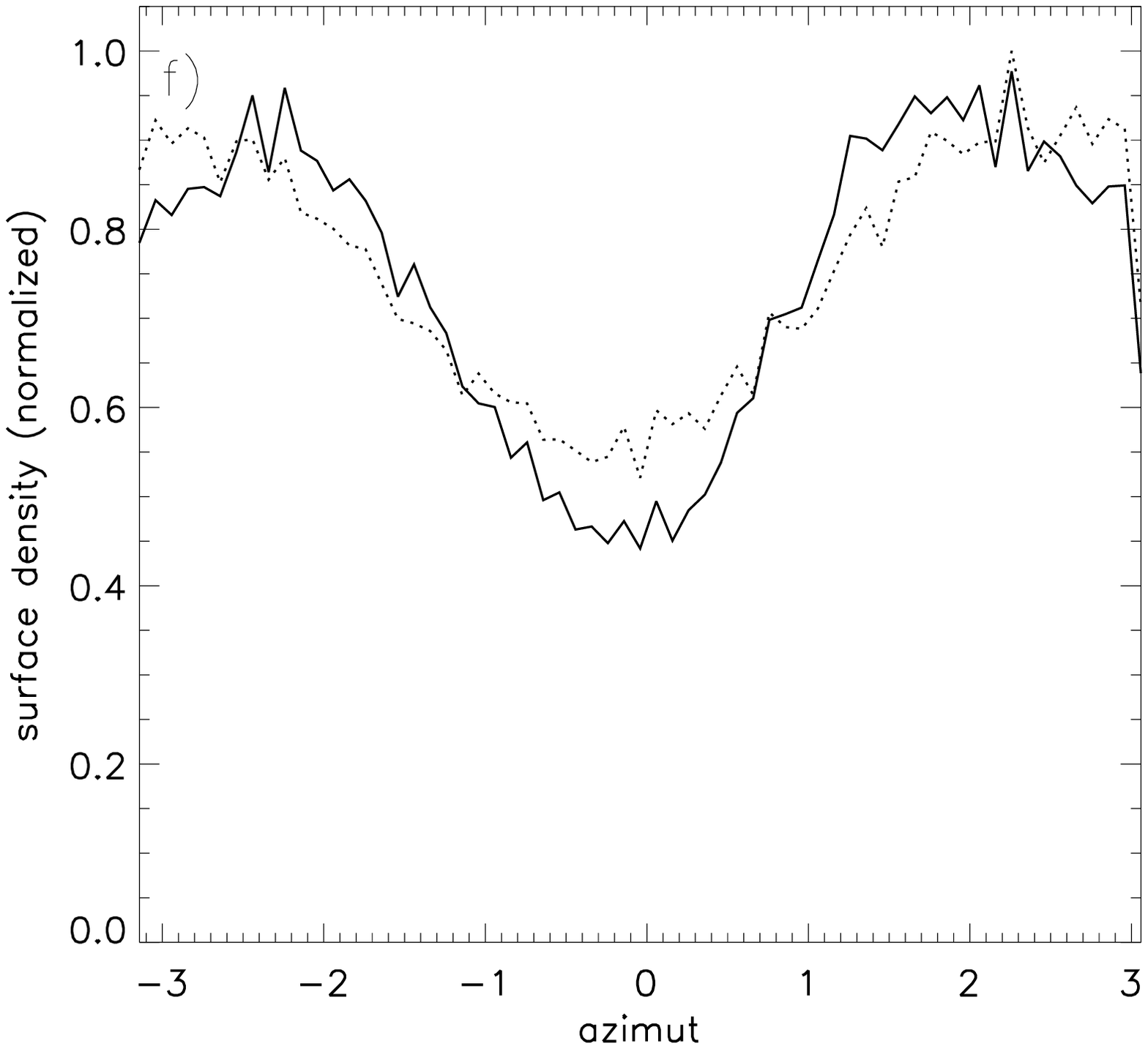}
}
\makebox[\textwidth]{
\includegraphics[scale=0.365]{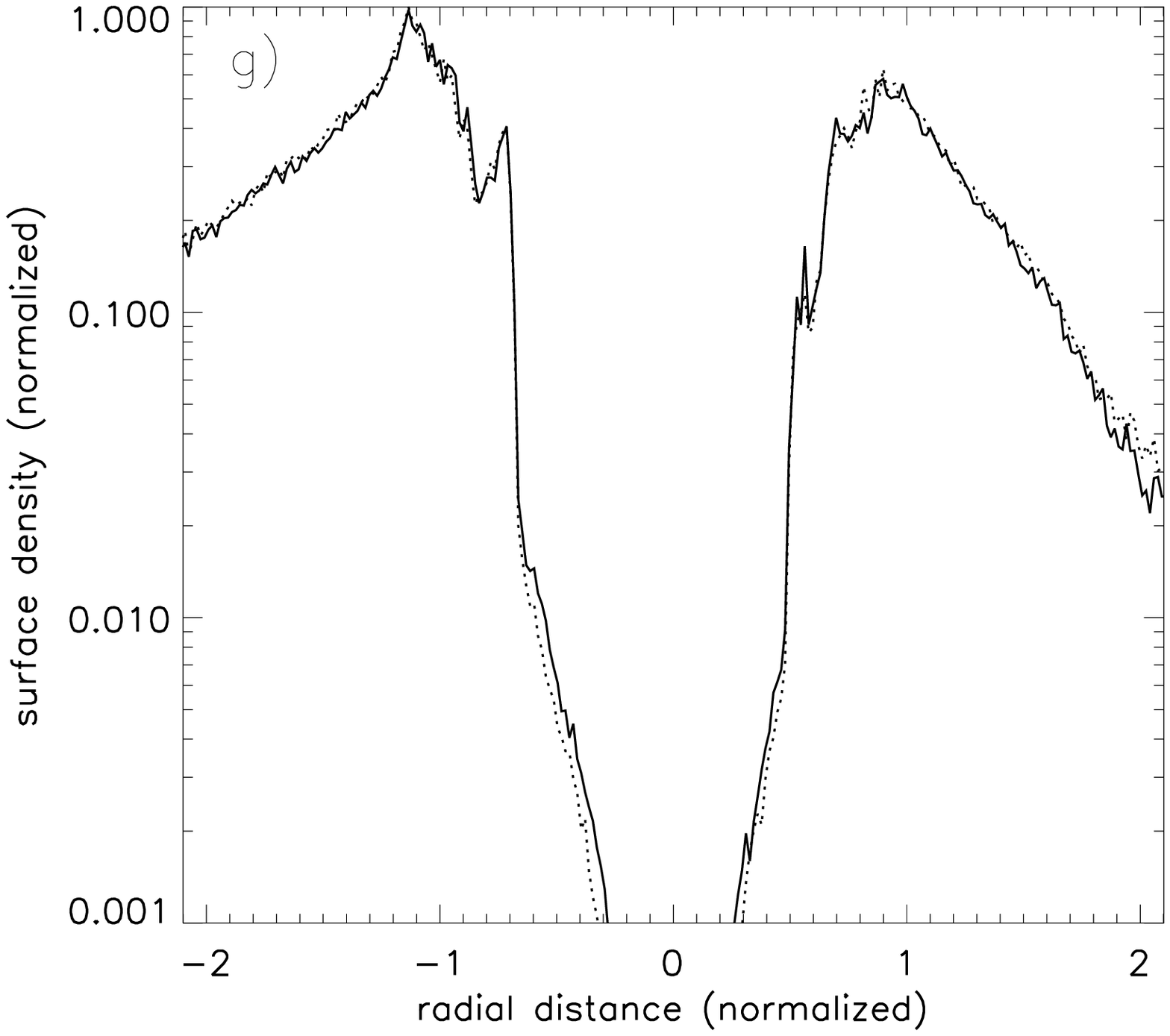}
\includegraphics[scale=0.365]{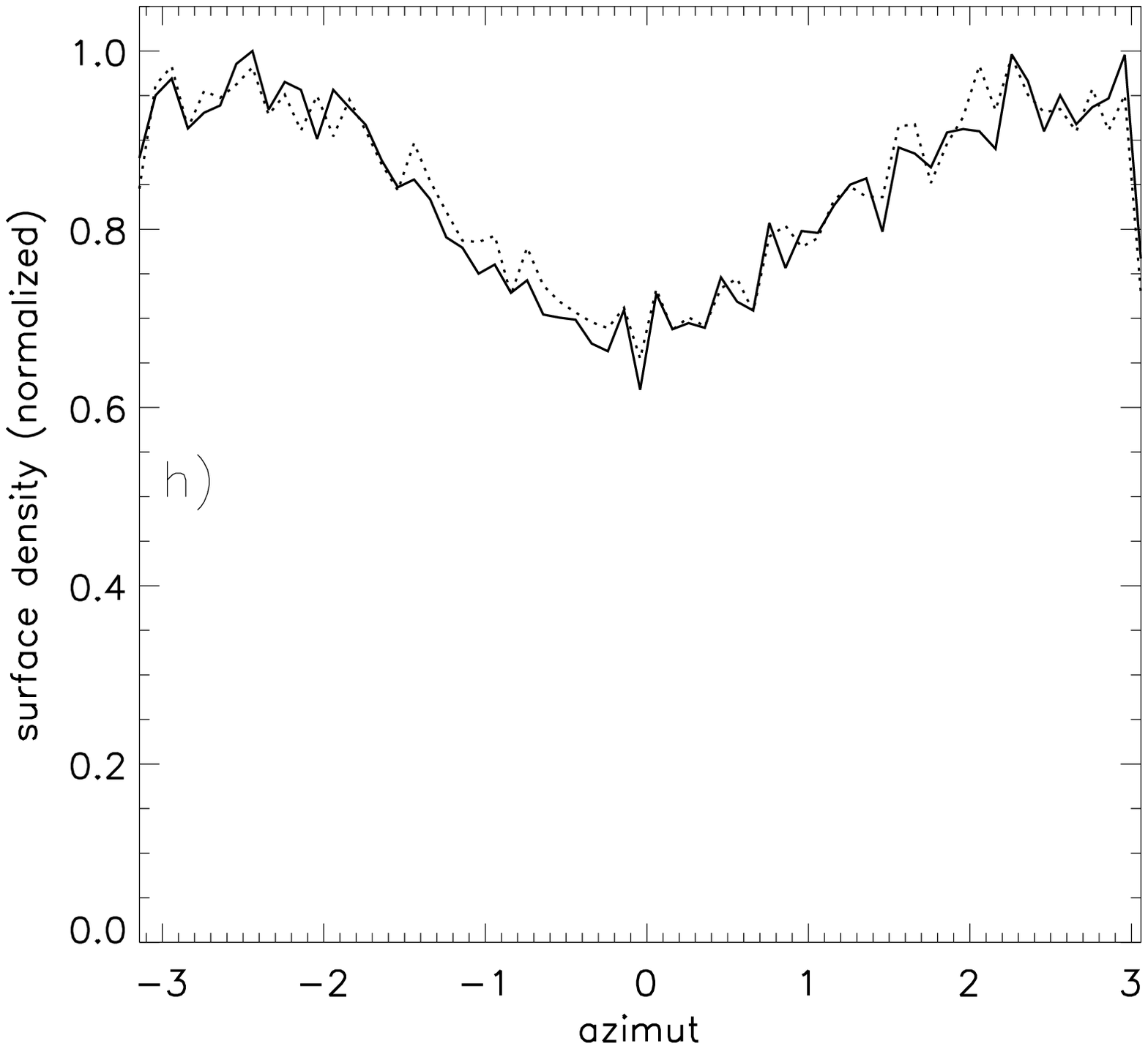}
}
\caption[]{Steady-state surface density profiles for the same (from top to bottom) $e_b=0$, $e_b=0.2$, $e_b=0.5$ and $e_b=0.75$ cases as in Fig.\ref{final}. \emph{Left panels}: radial profiles, along the binary semi-major axis direction, at periastron (full line) and apoastron (dotted line) passages of the companion star. \emph{Right panels}: azimutal profiles, for a $\Delta r=0.1$ wide annulus centered around $r=1$ (in $r_{crit}$ units), at periastron (full line) and apoastron (dotted line) passages of the companion star.
}
\label{azirad}
\end{figure*}

Regarding the spatial structure we find that, for low $e_B$ values, precessing spiral features directed towards the companion's position develop from the main ring \footnote{These precessing spirals outside the parent body region should not be confused with the transient spirals that develop within the parent body ring and disappear once the initial conditions have been relaxed (see previous paragraph)}. These precessing spirals are the most pronounced for the $e_B=0$ case, being clearly visible for the whole binary orbit with no significant shape evolution, as is logically expected for this circular orbit case.
As $e_B$ increases, these spiral structures become fainter. For the $e_B=0.2$ case, they are only visible during half of the binary orbit (from periastron to apoastron), while they only briefly appear at periastron passages in the $e_B=0.5$ run. For the $e_B=0.75$ case no spiral is longer visible. 
This disappearance of the precessing spirals for high $e_B$ reflects the more general trend of the disc tending towards a time-invariant structure. In particular, no significant shape change is observed at different orbital phases of the binary. 
The disc assumes a fixed and strongly asymmetric shape, extending much further out in the apoastron direction than in the periastron one (see radial profiles in Fig.\ref{azirad}e and g).

Another important result is the presence of strong inhomogeneities in the main parent body ring itself. For low $e_B$ cases, they consist of a pronounced dip in the opposite direction of the companion's position (Fig.\ref{azirad}b), which precesses at the binary's angular velocity. In the moderately eccentric case ($e_B=0.2$), this dip is surrounded by two overdensitites, precessing at the same rate and symetrically positioned at $\pm 70-90 $ with respect to the companion's antipolar position (Fig.\ref{azirad}d). The density contrast around the density dip can reach up to $\sim 60\%$ (Fig.\ref{azirad}b). For the $e_B=0.2$ case, in addition to the precessing features, a \emph{time-invariant} structure appears, i.e., a dip in the binary orbit's periastron direction (Fig.\ref{azirad}d). As $e_B$ increases even more ($e_B \geq 0.5$), this time-invariant asymmetry gets more pronounced, while the precessing structures progressively fade away (Fig.\ref{azirad}f and h).

\subsection{Discussion}

The results of the previous section show that a companion star can significantly affect the shape of a circumprimary disc, inducing pronounced radial and azimutal asymmetries. Basically, two different regimes can be identified depending on the binary's eccentricity.
 
For low $e_B$ binaries, the disc's shape is time-varying and precessing with the binary's orbit. Strong asymmetries are observed, whose position is always related to that of the companion on its orbit. Strong asymmetries are also observed for highly eccentric binaries, but they are in this case almost time-invariant and the disc has roughly the same shape regardless of the location of the companion on its orbit. In these high $e_B$ cases, the main asymmetries are oriented with respect to the binary orbit's geometry, not to the actual position of the companion on the orbit.
Between these 2 regimes, there is an intermediate state (see the $e_B=0.5$ runs) where the disc has a globally invariant asymmetric structure with transient features close to the companion's periastron passages.

These fixed, precessing or transient features are caused by the combination of 3 distinct processes:
1) \emph{Collisional activity} within the PB ring, steadily producing vast quantities of small grains that carry a large fraction of the geometrical cross section, 
2) \emph{Radiation Pressure} on the these small fragments, which places them on high-eccentricity orbits, 
3) \emph{Gravitational perturbations} by the companion star that will eventually, but not immediatly, remove all grains beyond $r_{crit}$. 
Equilibrium between these 3 effects results in a steady-state where a large fraction of the small, high-$\beta$ fragments produced in the PB ring remain in the dynamically forbidden regions long enough to cause the appearance of pronounced structures, whose shape and evolution strongly depends on $e_B$. 

A careful examination of the variety of structures obtained for the different cases displayed in Fig.\ref{final} shows that the action of a companion star could be a potential explanation for some features that have been routinely observed in debris discs: spiral arms, azimutal inhomogeneties in ring-like structures, blobs, etc. It is worth stressing that all these structures are obtained at \emph{steady-state}, and thus do not appear during a brief specific period in the system's history, as would be the case for a fly-by with a passing star \citep{ardi05,rech09}.

Of course, this explanation should be taken with some caution. A first reason is that other mechanisms have already been found to be able to produce asymmetries and inhomogeneities in discs, the most commonly invoked being the gravitational pull of a planet \citep{kriv10}. Another important point is that, contrary to the planet-perturbation scenario for which the presence of an undetected planet can often be freely postulated given the observational constraints on the system, the presence or absence of a companion star is usually much better constrained. Of all the 25 \emph{resolved} debris discs (as of August 2011, see http://circumstellardisks.org/), only 3 are known to inhabit a confirmed binary system: HR4796A, HD181296 and $\zeta$ Ret (HD141569A is more a "transitional" disc than a bona fide debris disc in the sense of the definition given by \cite{lag00}). A detailed analysis of these systems exceeds the scope of the present numerical paper, as it would require exploring several additional free parameters, such as $\tau$ or the mass ration $\mu$. It will be undertaken in a forthcoming study.

\section{Limitations and Perspectives}

It should be pointed out that this code is only a first, albeit important, step towards a fully integrated dynamics + collisions model. Let us here list its main limitations and the possible improvements that can be expected.
\begin{itemize}
\item The most obvious limitation is that collisions are not modelled in a self-consistent way, but through an empirically defined collision destruction probability assigned to each particle, and that no feedback of the collisions on the dynamics is included. These two problematic issues can probably not be handled by an N-body approach, for which the present code is maybe the upper limit of what can be achieved, at least for the case of high-velocity fragment-producing collisions. For this case of fragmenting impacts, a fully integrated and self-consistent dynamics+collisions model is still out of reach today, and can probably only be achieved by the next generation of "hybrid" codes in the spirit of the pioneering model explored by \citet{grig07}.
\item Our procedure is suited for systems where a level of symmetry has been reached, which is that the location of the parent bodies is symmetric with respect to the perturber's orbit, i.e., the global shape of the PB disc is identical between 2 passages of the perturber at the same orbital position. Note that this symmetry is not required for each \emph{individual} PB's orbit, which would be impossible because of the different periodicities between PBs and the perturber. What is required is a global symmetry for orbital streamlines, providing that each parent body's orbit is populated by a large number of bodies distributed uniformly in mean anomaly. This assumption of a phasing between global PB locations and pertuber orbit might not be valid for all perturbed disc configurations, in particular for systems strongly shaped by mean motion resonances. This case, usually occuring for discs interacting with embedded planets, will be explored in a forthcoming study. Note however that, for the present case of a circumprimary disc in a binary, this global symmetry is \emph{not} assumed or postulated but is a verified behaviour of the PB disc, which always reaches this state after a transitory period of 100-300 binary periods (see Sec.\ref{resu}).
\item The equation governing collisional probability does not resolve vertical structure in the disc because it is based on the vertically-integrated optical depth. This restrics the current version of the model to 2 dimensional problems, but those have been by far the most widely investigated by all past studies. A 3-D version of the collisional probability prescription is however fully implementable, at the cost of an increase in the size of the saved density maps at the end of the parent body runs.
\item Our model assumes that all collisions take place in the region of the parent body belt. This is not a major limitation for the present case of highly collisional massive debris discs, but the model will not perform well for fainter, drag-dominated discs where collisions can occur far interior to the belt. Our code can thus not include significant drag effects in its current version.
\end{itemize}

\section{Conclusions}

We have developed a new code to study the spatial structure of dynamically perturbed debris discs (be it by planets or companion stars) taking into account the effect of collisions. It is especially aimed at quantifying the competing effects between steady collisional production of small grains, placed by radiation pressure in dynamically unstable regions, and removal by gravitational perturbations. The principle of our numerical method is the following: we assume that the system has reached steady state and we perform a series of distinct N-body runs, each corresponding to a different initial position of the perturber on its orbit, for which each particle has a collisional destruction probability depending on its size and location. For each run, all particle positions are recorded at regularly spaced time intervals. This database of "collisional" runs is then used to reconstruct the steady state profile of the disc at different orbital phases of the perturber.

To illustrate the potential of this numerical procedure, we have applied it to the case of a circumprimary debris disc in a binary. We show that the complex interplay between collisional activity, radiation pressure and gravitational perturbations can create pronounced structures inside and outside the dynamical stability region. Depending on the binary's eccentricity, two regimes can be distinguished. For low $e_B$, the disc's structure is time varying, with spiral arms forming in the dynamically "forbidden" region beyond $r_{crit}$ and precessing at the binary's angular velocity. For high $e_B$, spiral structures disappear, the disc adopts a time invariant structure, extends far outside the stability region in the binary's apoastron direction and has a pronounced material depletion, inside $r_{crit}$, in the periastron direction. 

Our model has the potential to be applied to many other configurations, in particular debris discs sculpted by planets, which will be the purpose of a future study.

\begin{acknowledgements}

We thank the reviewer, Christopher Stark, for very useful comments that helped improve the quality of the manuscript. We thank the French National Research Agency (ANR) for financial support through contract ANR-2010 BLAN-0505-01 (EXOZODI).

\end{acknowledgements}

{}
\clearpage

\end{document}